\documentstyle[a4,12pt,epsf]{article}
\begin{document}
\def\seCtion#1{\section{#1} \setcounter{equation}{0}}
\renewcommand\theequation{\ifnum\value{section}>0{\thesection.
\arabic{equation}}\fi}
\newcommand{\be}{\begin{equation}}
\newcommand{\ee}{\end{equation}}
\newcommand{\bea}{\begin{eqnarray}}
\newcommand{\eea}{\end{eqnarray}}
\newcommand{\nn}{\nonumber}
\newcommand{\gauge}{{\rm{gauge}}}
\newcommand{\fb}{{\rm{fb}}}
\newcommand{\gf}{{\rm{gf}}}
\newcommand{\V}{{\rm{V}}}
\newcommand{\Tr}{{\rm{Tr}}}
\date{}
\pagestyle{empty}
\begin{titlepage}
\begin{flushright}
CERN-TH/95-72\\
FTUAM-95/8
\end{flushright}
\quad\\
\vspace{1cm}
\begin{center}
{\LARGE{\bf Fermionic Dispersion Relations}}\\
\medskip
{\LARGE{\bf in the Standard Model }}\\
\medskip
{\LARGE{\bf at Finite Temperature }}\\
\vspace{1.5cm}
{\bf C.~Quimbay\footnote{On leave of absence from Dpto. de
F\'{\i}sica Te\'orica, Univ. Aut\'onoma de Madrid, Cantoblanco,
28049 Madrid, and Centro Internacional de F\'{\i}sica, Bogot\'a.}
and S.~Vargas-Castrill\'on\footnote{On leave of absence from Dpto.
de F\'{\i}sica Te\'orica, Univ. Aut\'onoma de Madrid, Cantoblanco,
28049 Madrid.}}\\
\vspace{0.2cm}
{\it CERN, TH Division\\
CH-1211, Geneva 23\\
Switzerland\\}

\vspace{2.5cm}

\begin{abstract}
We compute the one-loop dispersion relations at finite temperature
for quarks, charged leptons and neutrinos in the Minimal Standard
Model. The dispersion relations are calculated in two different
plasma situations: for a vacuum expectation value $\upsilon$ of the
Higgs field $\upsilon \neq 0$ (broken electroweak symmetry) and for
$\upsilon=0$ (unbroken electroweak symmetry). The flavour and
chiral non-degeneracy of the quasi-particle spectrum is studied.
Numerical results show that the thermal effective masses for
fermions in the broken phase have a smaller value than those in
the unbroken phase. The temperature dependence of the top quark
and electron neutrino thermal effective masses is also presented.
Gauge invariance of one-loop dispersion relations is studied.
\end{abstract}
\end{center}

\vspace{2.0cm}
$\,$\\
CERN-TH/95-72\\
April 1995
\end{titlepage}
\newpage
\pagestyle{plain}

\seCtion{Introduction}

\hspace{3.0mm}
The behaviour of the plasma in the high temperature regime plays a
fundamental role in the attempts to explain several puzzles in
cosmology, as for example the present baryon asymmetry of the
universe \cite{sharef,barref} in the electroweak scenario. When the
temperature of the universe $(T)$ was near the critical one of the
cosmological electroweak phase transition $(T_C)$, the dynamics of
the plasma was governed by the interactions of the Minimal Standard
Model (MSM). It is now well known that the interaction of a fermion
with the thermal background in a plasma modifies the poles of the
fermion propagator with respect to those at tree level. In a
pioneering work \cite{welref}, Weldon studied the modification of
the fermion dispersion relation, i.e. poles of the fermion
propagator, due to high temperature effects for a chirally invariant
gauge theory (for instance, QCD or QED with massless fermions).
In the same work, he showed that for a chirally invariant gauge
theory with parity violation, as is the case for the
$SU(2)_L \otimes U(1)_Y$ model, the dispersion relations for the
right- and left-handed massless fermions are independent. This
decoupling implies that the thermal effective fermion masses,
defined as the energy values for zero momentum, are different.

For a parity preserving gauge theory at finite temperature, the
fermion dispersion relation has two possible solutions.
These solutions (branches in the phase space), both of positive
energy, describe the propagation of the fermionic excitations of
the plasma (quasi-particles) through the thermal medium. In the
literature, these solutions are generally known as normal and
abnormal branches. Each point of a branch corresponds to the
energy and momentum values of a quasi-particle. The
quasi-particles described by the normal branch (quasi-fermions)
have the helicity equal to their chirality, while those
described by the abnormal branch (quasi-holes) have the
helicity opposite to their chirality. This was studied in detail
by Weldon for massless quarks in QCD \cite{wel1ref,wel2ref} and
by Pisarski for light massive fermions in QCD/QED \cite{pisref}.
The normal branch is so called because it becomes the ordinary
fermionic branch at large momentum (large mass), $k~(M_f)>gT$,
where $g$ stands for the coupling constant and $M_f$ for the
fermion mass. The abnormal branch owes its name to the fact that
it appears as an additional physical solution only if temperature
effects are taken into account. Quasi-particles reflect a
collective behaviour of the plasma system at low momenta
\cite{blaref}. The collective behaviour of the quasi-holes only
exists for $k<gT$, where their creation probabilities are
non-zero. Moreover, the QCD/QED quasi-holes are physically
possible for $M_f<gT$ and they disappear from the spectrum for
$M_f \simeq T$ \cite{pisref,petref}.

The theoretical expectation that a cosmological electroweak phase
transition \cite{linref} took place during the cooling of the
early universe is motivated by the existence of the electroweak
spontaneous symmetry breaking (ESSB) in the MSM and its
extensions. The idea of the electroweak phase transition in the
framework of the MSM allows the consideration of the two following
phases. For $T>T_C$, the gauge symmetry of the fermion fields
is $SU(3)_C \otimes SU(2)_L \otimes U(1)_Y$, $\upsilon=0$ and the
plasma scenario is the electroweak unbroken phase. For $T<T_C$,
the electroweak symmetry $SU(2)_L \otimes U(1)_Y$ is spontaneously
broken. The gauge symmetry of the fermion fields is
$SU(3)_C \otimes U(1)_{EM}$, $\upsilon \neq 0$, and the plasma
scenario is the electroweak broken phase.

Fermionic excitations are present in both electroweak phases
of the plasma. The spectrum of quasi-particles can be obtained
from the thermal loop contribution to the fermion self-energy.
The energy of the quasi-particles is of order $g_{s}T$.
Quasi-particles, during their propagation in the plasma, scatter
with the thermal medium and this phenomenon induces a damping
rate $\gamma$ \cite{wel3ref}. This $\gamma$ is proportional
to the imaginary part of the self-energy \cite{pis1ref}. The
energy of the quasi-particles compared to $\gamma$ is large
and it is thus possible to speak about the existence of
quasi-particles as coherent excited states. Quasi-particles have
a finite lifetime $\sim1/2\gamma$ and they can eventually turn
into a new state, which is out of phase with the initial one. If
the group velocity is $|{\vec v}|$, the mean free path is
$|{\vec v}|/2\gamma$. The quantum spatial coherence, i.e. the
phase relation between points separated by a distance
$\geq$ $|{\vec v}|/2\gamma$, is lost \cite{barref,hueref}.
Since we are interested in obtaining the branches which describe
the quasi-particles, only the real part of the one-loop
contribution to the self-energy will be calculated.

The main purpose of the present work is to compute exactly the
one-loop dispersion relations for the different fermion sectors
of the MSM for the following situations of the plasma: for the
electroweak broken phase and for the electroweak unbroken one. We
study numerically the behaviour of the normal and abnormal branches
taking into account all terms in the fermion self-energy\footnote
{By exact numerical computation of the fermion dispersion
relations, we mean that we have included the leading and non-leading
$T$ corrections and the fermionic and bosonic mass corrections.}.
The flavour and chiral non-degeneracy of the quasi-particle
spectrum is studied. Our attention is focused in the low-momentum
regime where collective phenomena develop. Since mass and $T$
corrections are considered numerically, we obtain a novel result,
namely that thermal effective masses for fermions have a lower
value in the broken phase than in the unbroken one. The existence
of a similar shift was observed in \cite{monref} where, through a
lattice calculation, it was found that thermal effective masses for
massive bosons are smaller than those for massless bosons. We also
confirm the particular behaviour of the branches in the broken
phase found in \cite{sharef,barref}: at variance with what happens
in the unbroken phase, an energy gap of order $M_f$ is obtained
between the two branches that now have hyperbolic form. This
behaviour is crucial for the mechanism of electroweak baryogenesis
produced at the scattering of quasi-particles off the boundary
created during the first order electroweak phase transition.

The $T$ dependence for the top quark dispersion relations
in the broken phase and for the neutrino dispersion relations in
both phases is also studied. We have found, for the top quark,
that the thermal effective mass for two of the four branches can
have negative value for $T \leq M_t$. In order to study the gauge
independence of the dispersion relations, we compare one-loop
computations in Feynman and Landau gauges. The dispersion relations
are only gauge independent at leading order.

In section 2, we define the fermion dispersion relations for the
different types of gauge theories involved in the calculations. In
section 3, the generic gauge boson and scalar diagrams of Fig.~2
are calculated in the Feynman gauge and the real part of the finite
temperature contribution to the fermion self-energy is given. In
section 4, the Lorentz-invariant functions for quarks, charged
leptons and neutrinos in the MSM are calculated and the
corresponding dispersion relations in both phases are derived.
In section 5, we study in detail the gauge invariance for the
one-loop dispersion relations for massless and massive quarks in
QCD. The results are extended to the MSM dispersion relations.
Numerical results of the dispersion relation computations are
presented in section 6. Our conclusions are summarized in section
7. Results about the calculation of the real part of the fermion
self-energy in the Landau gauge are presented in Appendix B. In
Appendix C, the Lorentz-invariant functions calculated in the
Landau gauge are also presented.

\seCtion{Some useful definitions}

\hspace{3.0mm}
The fermionic dispersion relations, for the different cases in which
the chiral and parity symmetry are preserved or violated by a
gauge theory, are presented in this section. The fermionic spectrum
of the quasi-particles in a plasma is described by the branches
obtained from the fermion dispersion relation, which is derived from
the real part of the thermal contribution to the self-energy. The
fermion self-energy is written as a linear combination of the
matrices 1, $K{\hspace{-3.1mm}\slash}$, $u{\hspace{-2.2mm}\slash}$
and $K{\hspace{-3.1mm}\slash}u{\hspace{-2.2mm}\slash}$,
where $u^{\mu}$ is the four-velocity of the plasma, with
$u_{\mu}u^{\mu}=1$, and $K^{\mu}$ is the fermion momentum. In the
one-loop calculation,
$K{\hspace{-3.1mm}\slash} u{\hspace{-2.2mm}\slash}$ terms are not
generated and the real part of the thermal contribution to the
fermion self-energy, for a gauge theory with parity and chirality
violations, has the following form \cite{welref}:
\be \mbox{Re}\,\Sigma'(K)=- K{\hspace{-3.1mm}\slash}(a_{L}L +a_{R}R)-
u{\hspace{-2.2mm}\slash}(b_{L}L +b_{R}R)-
M_{I}(c_{L}L +c_{R}R)\label{petit},\ee
where $L\equiv\frac{1}{2}(1-\gamma_5)$ and
$R\equiv\frac{1}{2}(1+\gamma_5)$ are the left- and right-handed
chiral projectors respectively, and $M_I$ is the fermion mass. The
functions $a_L$, $a_R$, $b_L$, $b_R$, $c_L$ and
$c_R$ are the chiral projections of the Lorentz-invariant
functions $a$, $b$, $c$ defined in the following way:
\be a=a_L L + a_R R,\ee
\be b=b_L L+ b_R R,\ee
\be c=c_L L + c_R R.\ee
These ~functions depend ~on the ~Lorentz scalars ~$\omega$ and $k$
defined by ~$\omega\equiv(K\cdot u)$
{}~and $k\equiv[(K\cdot u)^2-K^2]^{1/2}$.

The inverse fermion propagator is given by
\be S^{-1}(K)= {\cal L}{\hspace{-2.5mm}\slash}L+
\Re{\hspace{-2.5mm}\slash}R- \Lambda_{L} L- \Lambda_{R} R,\ee
where:
\be {\cal L}^{\mu}= ( 1 + a_L) K^{\mu} + b_L u^{\mu},\ee
\be {\Re}^{\mu}= ( 1 + a_R) K^{\mu} + b_R u^{\mu},\ee
\be \Lambda_{L}= M_I (1 - c_L),\ee
\be \Lambda_{R}= M_I (1 - c_R).\ee

The fermion propagator follows from the inversion of (2.~5),
\bea S=\frac{1}{D}\left[\left({\cal L}^2\Re{\hspace{-2.5mm}
\slash}- \Lambda_{L}\Lambda_{R}{\cal L}{\hspace{-2.5mm}\slash}
\right)L + \left(\Re^2{\cal L}{\hspace{-2.5mm}\slash}-
\Lambda_{L}\Lambda_{R}\Re{\hspace{-2.5mm}\slash}\right)R+
\Lambda_{R}L\left({\cal L}\cdot\Re- \Lambda_{L}\Lambda_{R}+
\frac{1}{2}\left[{\cal L}{\hspace{-2.5mm}\slash},
\Re{\hspace{-2.5mm}\slash}\right]\right)L\nn\right.\\
+\left.\Lambda_{L} R\left({\cal L}\cdot\Re -\Lambda_{L}\Lambda_{R}
+\frac{1}{2}\left[{\cal L}{\hspace{-2.5mm}\slash},
\Re{\hspace{-2.5mm}\slash}\right]\right)R\right].\quad\quad\eea

The poles of the propagator correspond to values of $\omega$ and
$k$ for which the determinant $D$ in (2.~10) vanishes
\cite{welref},
\be D(\omega,k)={\cal L}^2 {\Re}^2 -
2\Lambda_{L}\Lambda_{R}({\cal L}\cdot\Re)+
(\Lambda_{L}\Lambda_{R})^2=0. \ee

In the rest frame of the plasma $u=(1,\vec 0)$, Eq.~(2.~11)
leads to the fermion dispersion relation for a gauge theory with
chiral and parity violations (for instance, the MSM after the
ESSB):
\bea \left\{\left[ \omega(1+a_L)+b_L \right]^2-
k^2\left[ 1+a_L \right]^2 \right\}
\left\{\left[ \omega(1+a_R)+b_R \right]^2-
k^2\left[ 1+a_R \right]^2 \right\}-\quad\quad\quad\quad\nn\\
2M_I^2(1-c_L)(1-c_R)\left\{({\omega}^2-
k^2)(1+a_L)\left(1+a_R\right)+
\omega[(1+a_L)b_R+ (1+a_R)b_L]+ b_L b_R \right\}\nn\\
+M_I^4 (1-c_L)^2 (1-c_R )^2=0.\quad\quad\quad\quad
\quad\quad\quad\quad\quad\quad\quad\quad\quad\quad\eea
Equation (2.~12) describes in a coupled way the behaviour of
the left and right components. This equation is of fourth order
in $\omega$, so that for each value of $k$ there are four
solutions for $\omega$. Therefore four branches (two normal and
two abnormal) in the phase space are obtained from the dispersion
relation. It is not possible to associate a specific chirality
with the quasi-particles because the chiral symmetry is broken.

\vspace{5.0mm}
For massless fermions, the chiral symmetry is preserved by the
theory. The functions $\Lambda_{L}$ and $\Lambda_{R}$ in (2.~11)
are equal to zero and the poles of the propagator are found from
${\cal L}^2 {\Re}^2=0$. Thus, for a chirally invariant theory with
parity violation, as is the case of the MSM before the ESSB, the
dispersion relations are given by
\be [\omega(1+a_L)+b_L]^2- k^2[1+a_L]^2=0, \ee
\be [\omega(1+a_R)+b_R]^2- k^2[1+a_R]^2=0. \ee
Left- and right-handed components now obey decoupled relations.
Equations (2.~13) and (2.~14) are of second order in $\omega$;
as a consequence, for each momentum value $k$ there are two values
of $\omega$ re-solving each equation. For each fermion chirality,
two branches (one normal and one abnormal) are solutions of the
dispersion relation. Four branches describe the fermionic
quasi-particles of the plasma. The quasi-particles have a specific
chirality, in contrast with what happens in the chiral violation
case.

\vspace{5.0mm}
For a parity-symmetric gauge theory, the projections of the
Lorentz-invariant functions over the left- and right-handed
components coincide. Setting $a_L=a_R=a$, $b_L=b_R=b$ and
$c_L=c_R=c$ in (2.~12), the dispersion relation for a
parity-invariant gauge theory with chirality violation (for
instance, QCD with massive quarks) is obtained:
\be [\omega(1+a)+b]^2 -k^2[1+a]^2 -M_I^2[1-c]^2=0. \ee
The left- and right-handed components of the dispersion relation
are superimposed in this case, and the fermionic excitations are
only described by two branches.

\vspace{5.0mm}
If chiral symmetry is preserved by a parity-invariant gauge
theory, as is the case of QCD with massless quarks, the
dispersion relation is given by (2.~15), with $M_I=0$,
\be [\omega(1+a)+b]^2 -k^2[1+a]^2 =0.\ee

\seCtion{Real part of the fermion self-energy}

\hspace{3.0mm}
The finite temperature contribution of the one-loop fermion
self-energy is calculated in this section. The MSM Feynman diagrams
that contribute to the one-loop fermion self-energy have the form
of the generic diagrams of Figs.~1 and 2. These generic diagrams
are calculated, in the Feynman gauge, using the real time formalism
of thermal field theory \cite{jacref}. It is straightforward to
see that tadpole diagrams represented by Fig.~1 only contribute to
the imaginary part of the self-energy, as is shown in Appendix A.
The diagrams that contribute to the real part of the self-energy
are represented by the generic diagrams of Fig.~2. To calculate
the poles of the fermion propagator, we are only interested in the
contribution to the real part of the fermion self-energy from these
diagrams. The same calculation is also performed in the Landau
gauge (Appendix B).

Finite temperature Feynman rules for vertices are the same as
those in the MSM at $T=0$, while the thermal propagators for massive
fermions $S(p)$, massive gauge bosons $D_{\mu \nu}(p)$ and massive
scalar bosons $D(p)$ are given by
\be S(p)= ( p{\hspace{-1.9mm}\slash}+ M_1)
\left[\frac{1}{p^2-M_1^2+i \epsilon}+ i{\Gamma}_f(p)\right],\ee
\be D_{\mu \nu}(p)=-g_{\mu\nu}\left[\frac{1}{p^2-M_2^2+i\epsilon}
-i{\Gamma}_b(p)\right],\ee
\be D(p)=\frac{1}{p^2-M_2^2+i\epsilon}-i{\Gamma}_b(p),\ee
where the $T$ dependence  is introduced through the functions
${\Gamma}_f(p)$ and ${\Gamma}_b(p)$ defined by
\be {\Gamma}_f(p)\equiv 2\pi\delta(p^2-M_1^2)n_F (p\cdot u),\ee
\be {\Gamma}_b(p)\equiv 2\pi\delta(p^2-M_2^2)n_B (p\cdot u),\ee
where $n_F (p\cdot u)$ and $n_B (p\cdot u)$ have the following form:
\be n_F(p\cdot u)= [ e^{[|p\cdot u|/T]}+1]^{-1},\ee
\be n_B(p\cdot u)= [ e^{[|p\cdot u|/T]}-1]^{-1}.\ee

The contribution to the self-energy from the generic gauge boson
diagram of Fig.~2a is
\be \Sigma(K)= ig^2 C_c \int\frac{d^{4} p}{(2\pi)^4}
D_{\mu \nu}(p) {\gamma}^{\mu}S(p+K){\gamma}^{\nu},\ee
where $g$ is the coupling constant and $C_c$ is the quadratic
Casimir invariant of the re-\,presentation. Inserting expressions
(3.~1) and (3.~2) into (3.~8), it is possible to write:
$\Sigma(K)=\Sigma^0 (K)+\Sigma'(K)$, where $\Sigma^0 (K)$ is the
zero temperature contribution and $\Sigma'(K)$ is the finite
temperature contribution. Keeping only the real part
$\mbox{Re}\,\Sigma'(K)$ of the finite temperature contribution,
we obtain
\be \mbox{Re}\,\Sigma'(K) = 2g^2C_c \int\frac{d^{4}p}{(2\pi)^4}
(p{\hspace{-1.9mm}\slash}+K{\hspace{-3.1mm}\slash}-2M_1)
\left[\frac{{\Gamma}_b (p)}{(p+K)^2-M_1^2}-
\frac{{\Gamma}_f(p+K)}{p^2-M_2^2}\right],\ee
where the denominators are defined by their principal value.
Applying Weldon's procedure \cite{welref} and after performing the
integrations over $p_0$ and the two angular variables, the
following novel integrals over the modulus of the three-momentum
$p=|\vec p|$ are obtained:

\be \frac{1}{4}Tr(\mbox{Re}\,\Sigma')=-2g^2C_c M_1\int^\infty_0
\frac{dp}{8\pi^2}\frac{p}{k}\left[\frac{1}{\epsilon_2}L_2^{+}(p)
n_B(\epsilon_2)+\frac{1}{\epsilon_1}L_1^{+}(p)n_F(\epsilon_1)
\right],\quad\quad\ee
\bea \frac{1}{4}Tr(K{\hspace{-3.1mm}\slash}\,\mbox{Re}\,\Sigma')=
g^2C_c \int^\infty_0\frac{dp}{8\pi^2}\frac{p}{k}\left[\frac{1}
{\epsilon_2}\left(\frac{(\omega^2-k^2-\Delta)}{2}
L_2^{+}(p)+4pk\right)n_B(\epsilon_2)\nn\right.\quad\quad\\
\left.+\frac{1}{\epsilon_1}\left(\frac{(\omega^2-k^2-\Delta)}{2}
L_1^{+}(p)+4pk\right)n_F(\epsilon_1)\right],\quad\quad\eea
\be \frac{1}{4}Tr(u{\hspace{-2.2mm}\slash}\,\mbox{Re}\,\Sigma')=
g^2C_c \int^\infty_0\frac{dp}{8\pi^2}\frac{p}{k}
\left[\left(L_2^{-}(p)+\frac{\omega}{\epsilon_2}L_2^{+}(p)
\right)n_B(\epsilon_2)+ L_1^{-}(p)n_F(\epsilon_1)\right],\ee
with the logarithmic functions given by

\be L_1^{\pm}(p)=\log\left[\frac{\omega^2-k^2-
\Delta-2\epsilon_1\omega-2kp}
{\omega^2-k^2-\Delta-2\epsilon_1\omega+2kp}\right]\pm
\log\left[\frac{\omega^2-k^2-\Delta+2\epsilon_1\omega-2kp}
{\omega^2-k^2-\Delta+2\epsilon_1\omega+2kp}\right],\ee
\be L_2^{\pm}(p)=\log\left[\frac{\omega^2-k^2+
\Delta+2\epsilon_2\omega+2kp}
{\omega^2-k^2+\Delta+2\epsilon_2\omega-2kp}\right]\pm
\log\left[\frac{\omega^2-k^2+\Delta-2\epsilon_2\omega+2kp}
{\omega^2-k^2+\Delta-2\epsilon_2\omega-2kp}\right],\ee
where $k=|\vec k|$, $\epsilon_1=(p^2+M_1^2)^{1/2}$,
$\epsilon_2=(p^2+M_2^2)^{1/2}$, and $\Delta=M_2^2-M_1^2$. These
expressions for $M_1=M_2=0$ are identical to those derived by
Weldon \cite{welref} and for $M_1 \neq 0$ and $M_2=0$ to those
derived by Petitgirard \cite{petref}.

\vspace{2.0mm}
The contribution to the fermion self-energy from the generic
scalar boson diagram shown in Fig.~2b has the same form
as the gauge boson contribution. The factor $2g^2C_c$ in (3.~9)
is replaced by $|l^2|C'_{L,R}$, where $l$ is the Yukawa coupling
constant and $C'_{L,R}$ are given in terms of the matrices of
Clebsch-Gordan coefficients $\Gamma^i$. The generic scalar boson
contribution for left and right chiralities is proportional
to $(\Gamma^i{\Gamma^{i}}^\dagger)_{mm'}\equiv C'_L \delta_{mm'}$
and $({\Gamma^{i}}^{\dagger}\Gamma^i)_{nn'}\equiv C'_R \delta_{nn'}$
respectively \cite{welref}.

\vspace{2.0mm}
The integrals (3.~10)-(3.~12) are related to (2.~1) through

\be a(\omega,k)=\frac{1}{k^2}\left[\frac{1}{4}
Tr(K{\hspace{-3.1mm}\slash}\,\mbox{Re}\,\Sigma')-\omega\frac{1}{4}
Tr(u{\hspace{-2.2mm}\slash}\,\mbox{Re}\,\Sigma')\right],\ee
\be b(\omega,k)=\frac{1}{k^2}\left[(\omega^2-k^2)\frac{1}{4}
Tr(u{\hspace{-2.2mm}\slash}\,\mbox{Re}\,\Sigma')-\omega\frac{1}{4}
Tr(K{\hspace{-3.1mm}\slash}\,\mbox{Re}\,\Sigma')\right],\ee
\be c(\omega,k)=-\frac{1}{M_I}\frac{1}{4}
Tr(\mbox{Re}\,\Sigma'),\ee
where the Lorentz-invariant functions $a(\omega,k), b(\omega,k)$
and $c(\omega,k)$ can be written in terms of their chiral
projections, as was shown in (2.~2)-(2.~4), and $M_I$ is the mass
of the external fermion.

It is convenient to rewrite (3.~15)-(3.~17) in the following form:
\be a(\omega,k)=g^2C_c A(M_1,M_2),\ee
\be b(\omega,k)=g^2C_c B(M_1,M_2),\ee
\be c(\omega,k)=2g^2C_c \frac{M_1}{M_I} C(M_1,M_2),\ee
with
\bea A(M_1,M_2)=\frac{1}{k^2}\int^\infty_0\frac{dp}{8\pi^2}
\left(\left[-\frac{(\omega^2+k^2+\Delta)}{2k}\frac{p}
{\epsilon_2}L_2^{+}(p)-\frac{\omega p}{k}L_2^{-}(p)+\frac{4p^2}
{\epsilon_2}\right]n_B(\epsilon_2)\nn\right.\\
+\left.\left[\frac{(\omega^2-k^2-\Delta)}{2k}\frac{p}
{\epsilon_1}L_1^{+}(p)-\frac{\omega p}{k}L_1^{-}(p)+\frac{4p^2}
{\epsilon_1}\right]n_F(\epsilon_1)\right),\quad\eea
\bea B(M_1,M_2)=\frac{1}{k^2}\int^\infty_0\frac{dp}{8\pi^2}
\left(\left[\frac{(\omega^2-k^2)}{k}pL_2^{-}(p)+
\frac{(\omega^2-k^2+\Delta)}{2}\frac{\omega}{k}\frac{p}
{\epsilon_2}L_2^{+}(p)-\frac{4\omega p^2}{\epsilon_2}\right]
n_B(\epsilon_2)\nn\right.\\
\left.+\left[\frac{(\omega^2-k^2)}{k}pL_1^{-}(p)-
\frac{(\omega^2-k^2-\Delta)}{2}\frac{\omega}{k}\frac{p}
{\epsilon_1}L_1^{+}(p)-\frac{4\omega p^2}
{\epsilon_1}\right]n_F(\epsilon_1)\right),\quad\quad\eea
\be C(M_1,M_2)=\frac{1}{k}\int^\infty_0\frac{dp}{8\pi^2}
\left(\frac{p}{\epsilon_2}L_2^{+}(p)n_B(\epsilon_2)+
\frac{p}{\epsilon_1}L_1^{+}(p)n_F(\epsilon_1)\right).\ee

Since each one-loop diagram contribution to the real part of the
fermion self-energy is proportional to the generic gauge boson
contribution (3.~9), it is possible to write
\be a(\omega,k)=\sum_{n}f_n A(M_1,M_2),\ee
\be b(\omega,k)=\sum_{n}f_n B(M_1,M_2),\ee
\be c(\omega,k)=\sum_{n}g_n C(M_1,M_2),\ee
where the Lorentz-invariant functions are obtained by adding up
the contributions $n$ coming from each diagram involved in the
calculation; $M_1$ and $M_2$ refer to the internal fermion and
to the boson mass of the loop diagram, respectively.

\seCtion{Fermionic dispersion relations}

\hspace{3.0mm}
In this section, the one-loop dispersion relations at finite
temperature for quarks, charged leptons and neutrinos in the MSM
are derived. The chiral projections of the Lorentz-invariant
functions in the broken and in the unbroken phases are calculated
following the procedure explained in the last section. The
unbroken phase Lorentz-invariant functions calculated in the
Landau gauge are presented in Appendix C.

In the MSM, the $W$ and $Z$ gauge bosons and the fermions acquire
mass after the ESSB. These masses are written as functions
of $\upsilon$ in the form: $M_W^2=g^2\upsilon^2/4$,
$M_Z^2=(g^2 + g'^2)\upsilon^2/4$ and $M_F^2= l_F^2 \upsilon^2/2$,
where $g$ and $g'$ are the gauge coupling constants of $SU(2)_L$
and $U(1)_Y$ respectively and $l_F$ is the Yukawa coupling constant
of the fermion $F$ to the Higgs boson. The mass of the Higgs boson
is $M_H^2=\rho \upsilon^2/2$, where $\rho$ is a free parameter
of the Higgs potential. The thermal dependence of the boson and
fermion masses is given through $\upsilon$ by
$\upsilon(T) \simeq \upsilon_0 (1-T^2/T_C^2)^{1/2}$, where
$T_C \sim 100$ GeV is the critical temperature of the electroweak
phase transition and $\upsilon_0=246$ GeV is the vacuum expectation
value of the Higgs field at $T=0$. The latter expression is valid
for values of $T$ near and below $T_C$.

\subsection{Quark dispersion relations}

\hspace{3.0mm}
As we have already mentioned, the damping rate of the
quasi-particles in the plasma is proportional to the imaginary part
of the fermion self-energy. The damping rate is induced by the
incoherent scattering of the quasi-particles with the thermal
medium. The scattering of the quasi-particles associated to quarks
is generated by the strong and electroweak thermal interactions.
The most important contribution to the damping rate comes from QCD
thermal interactions. It has been calculated at leading order for
zero quasi-particle momentum in \cite{pis1ref} giving:
$\gamma_{QCD}\sim 0.15 g_s^2 T$, \,\ i.e. $\sim 19$ GeV at
$T=100$ GeV, where $g_s$ is the strong coupling constant.

Assuming that, in the vicinity of $k=0$, the damping rate is not
changed appreciably by the electroweak thermal interactions, the
finite lifetime of the quasi-particles is $\sim 1/40$ GeV$^{-1}$.
Since the group velocity is $\sim 1/3$, the mean free path is
$\sim 1/120$ GeV$^{-1}$. Thus, the quantum spatial coherence is
lost for points separated by a distance $\geq 1/120$ GeV$^{-1}$.
We show below that the electroweak corrections to the branches for
light quarks are of the order of the quark masses $M_I$. Thus, the
imaginary part of the QCD self-energy is much larger than the
electroweak contributions to the real part of the self-energy for
all quarks but the top. This means that coherent electroweak quantum
phenomena associated to quarks (for instance, CP violation) cannot
take place in the plasma because the lifetime of the
quasi-particles is very small with respect to the characteristic
time of the electroweak processes \cite{barref,hueref}. When a
quasi-particle scatters, the electroweak coherence is lost.

\subsubsection{Broken phase}

\hspace{3.0mm}
The thermal background interacting with quarks in the plasma is
constituted by massive quarks of three flavours with colour charge,
massive Higgs bosons, massive $W$ and $Z$ gauge bosons, photons
and gluons. Quarks of electric charge $2/3$ are called up
quarks and those of electric charge $-1/3$ down quarks.
The one-loop diagrams with photon, gluon, $Z$ gauge boson, neutral
Goldstone and Higgs boson, which contribute to the quark self-energy,
do not induce a change of flavour in the quark.  The diagrams with
an exchange of charged Goldstone and $W$ gauge bosons induce a
flavour change in the incoming quark $I$ to a different outgoing
quark $F$. In the latter contributions, the flavour $i$ of the
internal quark (inside the loop) runs over the up or down quark
flavours according to the type of the external quark (outside the
loop). The one-loop contribution from gauge boson and scalar boson
to the real part of the self-energy gives
\be a_{\stackrel{L}{R}}(\omega,k)_{IF}=[f A(M_I,0)+
f_{\stackrel{L}{R}}^{Z}A(M_I,M_Z)+f^{H} A(M_I,M_H)]\delta_{IF}+
\sum_{i}f_{\stackrel{L}{R}}^{W}A(M_i,M_W),\ee
\be b_{\stackrel{L}{R}}(\omega,k)_{IF}=[f B(M_I,0)+
f_{\stackrel{L}{R}}^{Z}B(M_I,M_Z)+f^{H} B(M_I,M_H)]\delta_{IF}+
\sum_{i}f_{\stackrel{L}{R}}^{W}B(M_i,M_W),\ee
\be c_{\stackrel{L}{R}}(\omega,k)_{IF}=[2f C(M_I,0)+
g^{Z}C(M_I,M_Z)-f^{H} C(M_I,M_H)]\delta_{IF}+
\sum_{i}g_{\stackrel{L}{R}}^{W}C(M_i,M_W).\ee
The integrals (3.~21)-(3.~23) will be evaluated with the appropriate
values of $M_1$ and $M_2$ for each contribution. The coefficients
$f_n$ and $g_n$ are:
\be f=\frac{4}{3}g_s^2 + g^2 Q_I^2 s_w^2,\ee
\be f_L^{Z}=\frac{g^2}{c_w^2}(T_I^3 - Q_I s_w^2)^2+
\frac{g^2\lambda_I^2}{8},\quad\quad
f_R^{Z}=\frac{g^2}{c_w^2}Q_I^2 s_w^4 +
\frac{g^2\lambda_I^2}{8},\ee
\be f^{H}=\frac{g^2\lambda_I^2}{8},\ee
\be f_L^{W}=\frac{g^2}{2}(1+
\frac{\lambda_i^2}{2})K_{Ii}^{+}K_{iF},\quad\quad\quad\quad
f_R^{W}=\frac{g^2\lambda_I\lambda_J}{4}K_{Ii}^{+}K_{iF},\ee
\be g^{Z}=-\frac{2g^2}{c_w^2}(T_I^3- Q_I s_w^2)Q_I s_w^2+
\frac{g^2\lambda_I^2}{8},\ee
\be g_L^{W}=\frac{g^2\lambda_i^2}{4}K_{Ii}^{+}K_{iF},
\quad\quad\quad\quad\quad\quad\quad
g_R^{W}=\frac{g^2\lambda_i^2\lambda_F}{4\lambda_I}
K_{Ii}^{+}K_{iF},\ee
where ${\it K}$ represents the ~Cabibbo-Kobayashi-Maskawa ~(CKM)
matrix, $\lambda_I$ is defined as $\lambda_I=M_I/M_W$, $g$ is the
weak coupling constant, $Q_I$ is the quark charge and $s_w$
($c_w$) is the sine (cosine) of the electroweak mixing angle
$\theta_w$. The Yukawa coupling constant $l_I$ is related to
$\lambda_I$ through $l_I=g \lambda_I/ \sqrt{2}$. The substitution of
the expressions (4.~1)-(4.~3) in (2.~12) results in a non-diagonal
matrix because of the presence of mixing in the CKM matrix. From the
quark dispersion relation obtained, the branches are computed. The
mass non-degeneracy of the quark spectrum and the different Yukawa
coupling constants for the distinct quark flavours introduce the
flavour non-degeneracy of the branches. The branches associated to
each quark flavour are different because of the presence of $M_I$,
$\lambda_I$, $Q_I$ and the CKM matrix in (4.~1)-(4.~3).

The spectrum of quasi-particles for massive quarks with only QCD
interactions was studied in \cite{pisref,petref}. The
quasi-particles are described by one normal and one abnormal branch.
If electroweak effects are introduced, as they are here, the
mentioned spectrum is modified in a complicated way. Although the
electroweak corrections are quantitatively small for all flavours but
the top, some important consequences are derived from these effects.
In particularly, as was explained in section 2, four branches are
present (two normal and two abnormal) instead of the two QCD ones.
The four branches for the $s$ and $c$ quarks are shown in
Figs.~3 and 4. Because the chiral symmetry of the theory with massive
quarks is broken, it is impossible to associate a specific chirality
to the branches. The top quark acquires a large mass during
the electroweak phase transition. We note a particular behaviour of
the top quark branches with the $T$ variation: a couple of branches
lowers down appreciably as $T$ is lowered ($M_t$ increases) as
is shown in Fig.~5, by means of the $T$ variation of the
thermal effective masses. To try to understand this special
behaviour of the top quark branches, we will first study the much
simpler QCD case.

It is very easy to check that (4.~1)-(4.~3) reduce to QCD functions
when $g$ is fixed to zero in the coefficients (4.~4)-(4.~9). That
is for $a_L=a_R=a$, $b_L=b_R=b$, $c_L=c_R=c$,
\be a(\omega,k)=f'A(M_I,0),\ee
\be b(\omega,k)=f'B(M_I,0),\ee
\be c(\omega,k)=2f'C(M_I,0),\ee
where $f'=\frac{4}{3} g_s^2$. Inserting these expressions in
(2.~15), the dispersion relation for massive quarks in QCD is
obtained. We have checked the validity of the analytic approximations
performed in \cite{petref} for light massive quarks $(M_I<<T)$. In
this case, the analytic forms of the normal and abnormal branches
for $k<<T$ are:
\be \omega = M_{+} + \frac{1}{M_{Lead}^2 + M_{+}^2}
\left[ \frac{M_{Lead}^2}{3M_{+}} + \frac{1}{2M_I}
\left( \frac{3M_{+}^2 - M_{Lead}^2}{3M_{+}}
\right)^2 \right] k^2 + {\cal O}(k^4),\ee
\be \omega = M_{-} + \frac{1}{M_{Lead}^2 + M_{-}^2}
\left[ \frac{M_{Lead}^2}{3M_{-}}-\frac{1}{2M_I}
\left( \frac{3M_{-}^2 - M_{Lead}^2}{3M_{-}}
\right)^2 \right] k^2 + {\cal O}(k^4),\ee
respectively, with $M_{+}=[(M_I^2 + 4M_{Lead}^2)^{1/2} + M_I ]/2$
and $M_{-}=[(M_I^2 + 4M_{Lead}^2)^{1/2} - M_I ]/2$, where
$M_{Lead}^2 = f' T^2/8$. If $k=0$ then $\omega = M_{+}$ for the
normal branch and $\omega = M_{-}$ for the abnormal branch.
These different values of the energy are interpreted as thermal
effective masses. For values of $k$ close to zero, the normal
branch is above $(M_I^2 + 4M_{Lead}^2)^{1/2}/2$ by $M_I/2$ while
the abnormal one is below that value by the same amount. The
separation between the branches is $M_I$ and this increases with
the quark mass. When the electroweak corrections are taken into
account, a similar effect is also present. An energy gap of
order $\sim M_I$ appears between the minimum and the maximum of
the hyperbolic branches (see Figs.~3 and 4). If the quark
mass is large, the energy gap is larger than the separation
between the thermal effective masses. As a consequence, the
normal and abnormal branches with the same thermal effective mass
appear closer. For some $T$, the thermal effective mass of smaller
energy can be negative, as is suggested by Fig.~5.

More works about the dispersion relation for massive quarks in
QCD can be found in the literature. Analytic expressions for
the dispersion relation were calculated, in the limit
$M_I^2+k^2>>M_{Lead}^2$, by Seibert \cite{seiref}. In the
limit $M_I^2>>M_{Lead}^2$ as was done in \cite{altref}, the
approximation for the dispersion relation
$\omega^2=k^2+M_I^2+M_{Lead}^2$ is valid. The properties of the
quark spectrum were studied by Lebedev and Smilga \cite{smiref}
considering a dynamical thermal mass for the quarks in the
tree-level matrix Green's Functions.

\subsubsection {Unbroken phase}

\hspace{3.0mm}
With the masses of the particles equal to zero, the expressions
(4.~1)-(4.~3) are reduced to the chiral projections of the
Lorentz-invariant functions in the unbroken phase:
\be a_L(\omega,k)_{IF}=l_{IF}A(0,0),\ee
\be b_L(\omega,k)_{IF}=l_{IF}B(0,0),\ee
\be a_R(\omega,k)_{IF}=r_{IF}A(0,0),\ee
\be b_R(\omega,k)_{IF}=r_{IF}B(0,0),\ee
where the coefficients $l_{IF}$ and $r_{IF}$ are given by
\be l_{IF}=8\left\{\left(\frac{2\pi\alpha_S}{3}
+\frac{3\pi\alpha_W}{8}\left[1+\frac{\tan^2\theta_W}{27}
+\frac{1}{3}\lambda_I^2\right]\right)\delta_{IF}
+\frac{\pi\alpha_W}{8} K_{Ii}^{+}\lambda_i^2 K_{iF}\right\},\ee
\be r_{IF}=8\left\{\frac{2\pi\alpha_S}{3}+\frac{\pi\alpha_W}{2}
\left[Q_I^2\tan^2 \theta_W+\frac{1}{2}\lambda_I^2\right]\right\}
\delta_{IF}.\ee
The branches for massless quarks are computed from the dispersion
relation obtained by substituting (4.~15)-(4.~18) in (2.~13)
and (2.~14). Two normal branches (one left-handed and one
right-handed) and two abnormal branches (one left-handed and
one right-handed) describe quasi-quarks and quasi-holes
respectively. The different Yukawa coupling constants for the
distinct quark flavours introduce the flavour non-degeneracy of
the branches. In Figs.~3 and 4, the $s$ and $c$ quark
branches are presented.

Calculating for $\omega,k \ll T$ the leading ${\cal O}(T^2)$
contribution to the expressions (4.~15)-(4.~18), it is possible
to observe analytically the chiral non-degeneracy of the
quasi-particle spectrum. The integrals $A(0,0)$ and $B(0,0)$ in
this limit allow us to write
\be a_L(\omega,k)_{IF}=\frac{1}{8}\frac{l_{IF}T^2}{k^2}
\left[1-F\left(\frac{\omega}{2k}\right)\right],\ee
\be b_L(\omega,k)_{IF}=-\frac{1}{8}\frac{l_{IF}T^2}{k}
\left[\frac{\omega}{k}+ (\frac{k}{w}-\frac{w}{k})F\left(
\frac{\omega}{2k}\right)\right],\ee
\be a_R(\omega,k)_{IF}=\frac{1}{8}\frac{r_{IF}T^2}{k^2}
\left[1-F\left( \frac{\omega}{2k} \right)\right],\ee
\be b_R(\omega,k)_{IF}=-\frac{1}{8}\frac{r_{IF}T^2}{k}
\left[\frac{\omega}{k}+ (\frac{k}{w}-\frac{w}{k})F
\left(\frac{\omega}{2k}\right)\right],\ee
where $F(x)$ is defined by:
\be F(x)=\frac{x}{2}\left[\log\left(\frac{x+1}{x-1}
\right)\right].\ee
In this limit, the quark self-energy has been calculated
for QCD in \cite{welref,kliref} and generalized for the MSM in
\cite{sharef}. The following analytical expressions are derived
by replacing (4.~21)-(4.~24) in (2.~13) and (2.~14):
\be \omega_{L,R}-k=\frac{M_{L,R}^2}{k}\left[1+\left[1-
\frac{\omega_{L,R}}{k}\right]\frac{1}{2}
\log\left(\frac{\omega_{L,R}+k}{\omega_{L,R}-k}\right)\right],\ee
where we have defined $M_L^2=l_{IJ}T^2/8$ and $M_R^2=r_{IJ}T^2/8$.
The normal and abnormal branches for small momentum
$(k << M_{L,R})$ are given by
\be \omega_{L,R}(k)=M_{L,R}+\frac{k}{3}+\frac{k^2}{3M_{L,R}}
+{\cal O}(k^3),\ee
\be \omega_{L,R}(k)=M_{L,R}-\frac{k}{3}+\frac{k^2}{3M_{L,R}}
+{\cal O}(k^3),\ee
respectively. For $k=0$, $\omega_{L,R}=M_{L,R}$ and then
$M_{L,R}$ are interpreted as thermal effective masses. For each
chirality, one normal and one abnormal branch describe the
quasi-particle spectrum. For high momentum $(k >> M_{L,R})$, the
branches are approximated by
\be \omega_{L,R}(k)= k + \frac{M_{L,R}^2}{k}- \frac{M_{L,R}^4}{2k^3}
\log\left(\frac{2k^2}{M_{L,R}^2}\right) + ...\ee
\be \omega_{L,R}(k)= k + 2k \cdot \exp\left[-\frac{2k^2}{M_{L,R}^2}
\right] + ...\ee
For $k >> M_{L,R}$ the branches become the ordinary fermionic
branch. The leading ${\cal O}(g_{s}^2T^2)$ expressions
(4.~21)-(4.~24) are gauge independent and they give place to the
gauge invariant dispersion relations (4.~27)-(4.~30) \cite{welref}.
Supposing that the propagation of quasi-particles is along the z
axis, it is possible to see that their group velocity is of
$\pm 1/3$. For $k_z$ and a given flavour, quasi-particles of a
specific chirality have two possible helicities. The group velocity
is given by the eigenvalue of
$\simeq 1/3\gamma_5\sigma_z = 1/3\gamma_5\sigma_z(\hat{k}_z)^2$,
i.e. $1/3\chi h\hat{k}_z$, where $h$ is twice the helicity, $\chi$
the chirality and $\hat{k}_z=k_z/|k_z|$ \cite{barref}. Consequently,
the group velocity for quasi-fermions has the same sign as the
momentum when chirality equals helicity. For quasi-holes, the group
velocity and the momentum have opposite signs. These fermionic
states have flipped chirality/helicity and at zero $T$ they have
negative energy. They become physical at finite $T$ due to a shift
in energy of the branch.

Turning off the electroweak interactions in the Lorentz-invariant
functions (4.~15)-(4.~18), the QCD case for massless quarks is
recovered, with
\be a(\omega,k)=f'A(0,0),\ee
\be b(\omega,k)=f'B(0,0).\ee
The corresponding dispersion relations follow from the substitution
of (4.~31) and (4.~32) in (2.~16). In the limit \,\ $\omega,k<<T$ \,\
the branches are given by (4.~27) and (4.~28), with
$\omega_L=\omega_R=\omega$ and $M_L=M_R=M_{Lead}$ in accordance
with \cite{welref,wel2ref,kliref}.

\subsection{Charged-lepton dispersion relations}

\hspace{3.0mm}
The scattering between the quasi-particles associated to charged
leptons and the thermal medium is governed by the weak and
electromagnetic thermal interactions. It is expected for $k=0$,
in analogy with QCD calculations \cite{pis1ref}, that the damping
rate at leading order originated only from weak or electromagnetic
thermal interactions were $\gamma_{EW} \sim g^2 T$. The weak
and electromagnetic leading contributions of the real part of the
self-energy are both of order $gT$. This means that
the energy of the quasi-particles, of order $gT$, is large in
comparison with the damping rate, i.e. it contains one additional
power of $g$. Thus, it is possible to speak of the existence of
quasi-particles as coherent excited states. Coherent electroweak
quantum phenomena associated to charged leptons can take place in
the plasma because the lifetime of the quasi-particles is of the
order of the characteristic time of the electroweak processes.

\subsubsection {Broken phase}

\hspace{3.0mm}
The thermal background interacting with charged leptons in the
plasma is constituted by massive charged leptons of flavour $e$,
massless neutrinos of flavour $e$, massive Higgs bosons, massive
$W$ and $Z$ gauge bosons and massless photons. The diagrams
with charged Goldstone and $W$ bosons change from the external
charged lepton to the associated neutrino flavour inside the loop.
The one-loop contribution from scalar and gauge boson sectors to
the real part of self-energy allow us to write
\be a_{\stackrel{L}{R}}(\omega,k)= dA(M_e,0)+
d_{\stackrel{L}{R}}^{Z}A(M_e,M_Z)+
d^{H} A(M_e,M_H)+ d_{\stackrel{L}{R}}^{W}A(0,M_W),\ee
\be b_{\stackrel{L}{R}}(\omega,k)= dB(M_e,0)+
d_{\stackrel{L}{R}}^{Z}B(M_e,M_Z)+
d^{H} B(M_e,M_H)+ d_{\stackrel{L}{R}}^{W}B(0,M_W),\ee
\be c_L(\omega,k)=c_R(\omega,k)=2dC(M_e,0)+ y^{Z}C(M_e,M_Z)-
d^{H}C(M_e,M_H),\ee
with the coefficients $d$ and $y$ given by
\be \quad\quad\quad\quad d= g^2 s_w^2, \ee
\be d_L^{Z}=\frac{g^2}{c_w^2}(\frac{1}{2}-s_w^2)^2+
\frac{g^2\lambda_e^2}{8},\quad\quad\quad\quad
d_R^{Z}=\frac{g^2}{c_w^2} s_w^4 + \frac{g^2\lambda_e^2}{8},\ee
\be \quad\quad\quad\quad  d^{H}= \frac{g^2\lambda_e^2}{8},\ee
\be \quad\quad\quad\quad
d_L^{W}= \frac{g^2}{2}, \quad\quad\quad\quad
d_R^{W}= \frac{g^2\lambda_e^2}{4},\ee
\be \quad\quad\quad\quad
y^{Z}=-\frac{2g^2}{c_w^2}(\frac{1}{2}-s_w^2)s_w^2+
\frac{g^2\lambda_e^2}{8},\ee
where $\lambda_e=M_e/M_W$. Substituting the functions (4.~33)-(4.~35)
in (2.~12), the branches are computed from the obtained dispersion
relation. Because the chiral symmetry in this case is broken, it is
impossible to associate a specific chirality to the branches.
Flavour non-degeneracy exists since the mass non-degeneracy is
present in (4.~33)-(4.~35).

The spectrum of quasi-particles for charged leptons in QED was
studied in \cite{becref}. For this case, the quasi-particles are
described by one normal and one abnormal branch, and left- and
right-handed branches are superposed. The breaking of the parity
symmetry by the weak interactions causes the doubling of the
QED branches. Thus, for charged leptons in the MSM, the
quasi-particles are described by two normal and two abnormal
branches, as is shown in Fig.~6 for the $\tau$ lepton. As in the
quark case in the broken phase, an energy gap between the hyperbolic
branches appears.

Following a procedure analogous to the one performed for massive
quarks in QCD, the Lorentz-invariant functions for massive
electrons in the QED case are obtained by turning off the weak
interactions into the functions (4.~33)-(4.~35). The expressions
derived are similar to expressions (4.~10)-(4.~12), with $f'$
changed to $d$ and $M_I$ to $M_e$. The analytic form of the two
branches, for $k<<T$, is given by (4.~13) and (4.~14).

Some aspects of the dispersion relation for massive electrons in
QED have been studied at finite temperature, considering an
effective chemical potential in \cite{levref,churef}. The spectrum
of quasi-particles in a hot relativistic electron plasma was studied
in \cite{bayref}, where it is shown numerically that the full
spectral weight is not gauge independent. By comparing one-loop
computations in the Feynman and in the Coulomb gauges, it was shown
in \cite{bayref} that the location of well-defined poles in the
propagator is gauge independent at leading order since the poles
correspond to physical excitations, while the widths of the
quasi-particles are gauge dependent. The dispersion relation at
zero temperature has been studied in \cite{toiref}.

\subsubsection {Unbroken phase}

\hspace{3.0mm}
Expressions (4.~33)-(4.~35) with particle masses equal to zero reduce
to the chiral projections of the Lorentz-invariant functions in the
unbroken phase:
\be a_L(\omega,k)=tA(0,0),\ee
\be b_L(\omega,k)=tB(0,0),\ee
\be a_R(\omega,k)=uA(0,0),\ee
\be b_R(\omega,k)=uB(0,0),\ee
where the coefficients $t$ and $u$ are
\be t=\pi\alpha_W \left[ 2 + \frac{1}{c_w^2}+\lambda_e^2\right],\ee
\be u= 4\pi\alpha_W \left[\tan^2\theta_W +
\frac{\lambda_e^2}{2}\right].\ee
The branches for massless charged leptons are computed
from the dispersion relation derived by substituting (4.~41)-(4.~44)
in (2.~13) and (2.~14).  Two normal branches (one for each chirality)
and two abnormal branches (one for each chirality) describe
quasi-particles and quasi-holes respectively (see Fig.~6). The
quasi-particle spectrum is flavour non-degenerate, but, because
$\lambda_e^2$ in (4.~45) and (4.~46) is so small for all charged
lepton flavours, the differences between flavours are very small.
The type of quasi-particle characterized by the abnormal branch
has been called the ``plasmino'' by Braaten and Pisarski
\cite{bra1ref}. In the limit $\omega,k \ll T$, the analytical forms
obtained for the dispersion relation are similar to those obtained
for massless quarks. For $k << M_{L,R}$, the branches are given by
(4.~27) and (4.~28) with $M_L^2=tT^2/8$ and $M_R^2=uT^2/8$.

\subsection{Neutrino dispersion relations}

\hspace{3.0mm}
The damping rate of quasi-particles associated to neutrinos is
induced by the weak thermal interactions. In analogy with the
assumption in the charged-lepton case, the damping rate (leading
order) of neutrino quasi-particles for $k=0$ is
$\gamma_{EW} \sim g^2 T$. The weak leading contribution of the
real part of the self-energy is of order $g T$. The energy of
the quasi-particles, of order $g T$, compared to the damping rate
is large. The quasi-particles exist as coherent excited states.
Coherent weak quantum phenomena associated with neutrinos can take
place in the plasma because the lifetime of the quasi-particles
is of the order of the characteristic time of the weak processes.

\subsubsection {Broken phase}

\hspace{3.0mm}
The thermal background interacting with neutrinos in the plasma
is constituted by massless neutrinos of flavour $e$, massive
charged leptons of flavour $e$ and massive $W$ and $Z$ gauge bosons.
The one-loop contribution from scalar and gauge boson sectors to
the real part of the self-energy lead to
\be a_L(\omega,k)= v_Z A(0,M_Z)+v_W A(M_e,M_W),\ee
\be b_L(\omega,k)= v_Z B(0,M_Z)+v_W B(M_e,M_W),\ee
with the coefficients $v_Z$ and $v_W$ given by
\be v_Z= \frac{\pi\alpha_W}{c_w^2},\ee
\be v_W= \pi\alpha_W [2 + \lambda_e^2],\ee
where $\lambda_e=M_e/M_W$. The projection of the Lorentz-invariant
functions is only over the left-handed component. Substituting
(4.~47) and (4.~48) in (2.~13), the branches are computed from the
dispersion relation obtained. A normal and an abnormal branches of
left-handed chirality describe the quasi-particles as is shown
in Fig.~7. Flavour non-degeneracy exists from the mass
non-degeneracy present in (4.~47) and (4.~48), for different neutrino
flavours. Because $M_e<<M_W$ in the integrals of (4.~47) and
(4.~48), the differences for different flavours are very small.

The neutrino dispersion relation presented here was studied by
D'Olivo, Nieves and To-\,rres in \cite{nieref} for temperatures of
the plasma much lower than the $W$ and $Z$ masses. This
approximation allows to neglect the thermal dependence for the
gauge boson propagators. These authors performed the calculation
in a general gauge and they found that the neutrino self-energy
depends on the gauge parameter, but the dispersion relation is
independent at leading order in $g^2/M_{W}^2$. Several works,
previous to \cite{nieref}, on this subject can be found in the
literature. In \cite{rafref,enqref}, the same calculation was
performed but using the unitary gauge. The computation of the
neutrino thermal self-energy was considered in
\cite{nie1ref,palref}.

\subsubsection {Unbroken phase}

\hspace{3.0mm}
Expressions (4.~47) and (4.~48) with gauge boson masses equal to
zero reduce to the left-handed projection of the Lorentz-invariant
functions in the unbroken phase:
\be a_L(\omega,k)= tA(0,0),\ee
\be b_L(\omega,k)= tB(0,0),\ee
with the coefficient $t$ given by (4.~45).  Substituting
(4.~51) and (4.~52) in (2.~13), the branches are computed from the
dispersion relation obtained. A normal and an abnormal branch of
left-handed chirality describe the quasi-particles
(see Fig.~7). The quasi-particle spectrum is flavour
non-degenerate, but, since $\lambda_e^2$ is very small for all
neutrino flavours, the differences for the different flavours are
very small. The variation with the temperature of the thermal
effective masses for both phases is presented in Fig.~8.

The branches for left-handed neutrinos are the same as those for
left-handed charged leptons. This result reflects the presence of
the $SU(2)_L$ symmetry in the MSM before the ESSB. The neutrino
dispersion relation in the unbroken phase was considered with a
non-zero chemical potential in \cite{kimref}. The same authors in
a later work \cite{kim1ref} studied the dispersion relation
in the case where the charged leptons have a running mass obtained
using the renormalization group equations.

\seCtion{Remarks about gauge invariance}

\hspace{3.0mm}
The gauge invariance of the one-loop dispersion relations
presented in this work is studied in this section. The numerical
results for the dispersion relations have been obtained
considering all the terms in the Lorentz-invariant functions
(3.~18)-(3.~20). This fact implies the introduction of a small
gauge dependence originated from the non-leading order terms of the
fermion self-energy \cite{kobref}. These, in turn, lead to
sub-leading contributions to the dispersion relations. Indeed, by
confronting QCD computations in the Feynman and Landau gauges, we
show that the sub-leading contributions to the dispersion relations
are gauge dependent. We also study the effect of these
contributions in our numerical results.

In Appendix B, we calculate the finite temperature contribution
to the one-loop real part of the fermion self-energy in the Landau
gauge $\mbox{Re}\,\Sigma'^{\pounds}(K)$. We show in (B.~4) that it
is possible to write:
$\mbox{Re}\,\Sigma'^{\pounds}(K)= \mbox{Re}\,\Sigma'(K) +
\mbox{Re}\,\delta\Sigma'(K)$, where $\mbox{Re}\,\Sigma'(K)$ is
the real part of the thermal contribution to the self-energy in the
Feynman gauge (3.~8) and $\mbox{Re}\,\delta\Sigma'(K)$ is the real
part of the gauge dependent thermal contribution of
$\mbox{Re}\,\Sigma'^{\pounds}(K)$ (B.~5). The Ward identities ensure
that the gauge dependent contribution disappears if the external
leg is projected on the zeroth order pole \cite{kobref}. This can
easily be seen because $\mbox{Re}\,\delta\Sigma'(K)$ can be
written as proportional to $K{\hspace{-3.1mm}\slash}-M_q$. In
QCD at finite temperature, for instance, the shift of the quark
propagator pole position with respect to the pole position at zero
temperature is of order $g_{s}T$. This means that the argument of
the integral in (B.~5) is proportional to $g_{s}T$, for
this case, because this integral is proportional to
$K{\hspace{-3.1mm}\slash}-M_q$. Consequently, the
$\mbox{Re}\,\delta\Sigma'(K)$ contribution to the dispersion
relation is of a higher order in $g_s$ with respect to the
$\mbox{Re}\,\Sigma'(K)$ contribution.

For example, in the QCD case for massless quarks,
$\mbox{Re}\,\delta\Sigma'(K)$ vanishes on the zeroth order
pole corresponding to $\omega = 0$, for $k=0$. So,
$\mbox{Re}\,\delta\Sigma'(K)$ contribution is a non-leading order
in $\omega/T$. These non-leading order contributions lead to
a sub-leading contributions to the thermal effective mass. This
fact can be guessed because the leg projection on the first order
pole means $\omega/T \sim g_s$. This conclusion is in
accordance with \cite{kobref}, where it is formally shown that
non-leading contributions to the quark self-energy allow gauge
dependent sub-leading modifications to the dispersion relation.

Let us now show in an explicit calculation that the non-leading
contributions to the one-loop QCD self-energy are sub-leading in
the dispersion relations. By comparing one-loop computations in
the Feynman and in the Landau gauge for massless and for massive
quarks, we will show that indeed the sub-leading contributions to
the dispersion relations are gauge dependent. First, we consider
massless quarks in QCD. The Lorentz-invariant functions in the
Feynman gauge (3.~18) and (3.~19) for this case are
\be a^{Q}(\omega,k) = \frac{M_{Lead}^2}{k^2}\left[1 -
F\left(\frac{\omega}{2k}\right)\right] + \frac{A^2}{k^3}
\left[\omega^2 I_1 + k^2 I_2 - 2\omega I_3 \right], \ee
\be b^{Q}(\omega,k) = \frac{M_{Lead}^2}{k}\left[\frac{\omega}{k}+
(\frac{k}{\omega}-\frac{\omega}{k})F\left(\frac{\omega}{2k}
\right)\right] + 2A^2\frac{\omega^2-k^2}{k^3}
\left[I_3 - \frac{\omega}{2} I_1\right], \ee
where $M_{Lead}^2= g_{s}^2T^2/6$, $A^2=g_{s}^2/12\pi^2$ and $g_s$
stands for the strong coupling constant. The function $F(x)$ is
defined by (4.~25) and the integrals $I_1$, $I_2$, $I_3$ are given
by
\be I_1(\omega,k,T)=\int^\infty_0 dp~L_I^{+}(p)
\left[n_{B}(p) + n_{F}(p)\right], \ee
\be I_2(\omega,k,T)=\int^\infty_0 dp~L_I^{+}(p)
\left[n_{B}(p) - n_{F}(p)\right], \ee
\be I_3(\omega,k,T)=\int^\infty_0 dp~p L_I^{+}(p)
\left[n_{B}(p)+ n_{F}(p)\right], \ee
where $n_{B}(p)$ and $n_{F}(p)$ are defined by (3.~6) and (3.~7)
respectively. The logarithmic functions $L_I^{\pm}(p)$ are
\be L_I^{\pm}(p) = \log \left[\frac{2p-(\omega-k)}
{2p-(\omega+k)}\right] \pm \log \left[
\frac{2p+(\omega+k)}{2p+(\omega-k)}\right].\ee

Substituting the expressions (5.~1) and (5.~2) into (2.~16), it
follows that
\be k^2=M_{Lead}^2\left(\frac{k}{\omega -k}-\frac{1}{2}\log
\left[\frac{\omega +k}{\omega -k}\right] \right)- A^2
\left(\omega I_1 - k I_2 -2 I_3\right). \ee
For $k \rightarrow 0$, the integrals $I_1$, $I_2$, $I_3$ can be
written as $I_{i}(\omega,k,T)=kF_{i}(\omega,T)+ {\cal O}(k^3)$,
where $i=1,2,3$. The reason of this is that (5.~3)-(5.~5) are
antisymmetric for the change $k \rightarrow -k$. It was not
possible for us to obtain analytic expressions for the integrals
(5.~3)-(5.~5). They were computed numerically for different values
of $\omega$, $k$ and $T$, in the limit $k \rightarrow 0$. By means
of a fit of the numerical results obtained, we have found an
analytical form for the integrals. Numerically we have proved that
$(\omega I_1 - 2 I_3 )= {\cal O}(k^3)$, i.e.
$(\omega F_1 - 2 F_3 )=0$. For this reason $(\omega I_1 - 2 I_3)$
vanishes in (5.~7), when we take the limit $k << \omega$,
$T>\omega$, divide by $k^2$ and set $k=0$. Then, we obtain the
following relation:
\be \omega^2\left[1-A^2 F_2(\omega,T)\right]=M_{Lead}^2, \ee
where\footnote{The analytical form found for $I_2$, for
$\omega \simeq T/2$, fits the numerical solutions with a precision
smaller than $0.01\%$.}
\be F_2(\omega,T)=2\gamma_{E} + 2\log \frac{\omega}{2T}, \ee
with $\gamma_{E}=0.5772$. We note that (5.~8) reduces to the
{}~leading thermal ~effective mass ~~$\omega=M_{Lead}$, if the
non-leading contributions given through $F_2(\omega,T)$ are
set to zero. The numerical solution of (5.~8) leads to the
thermal effective mass
\be M=M_{Lead}(1 + \rho ), \ee
where $\rho=0.02223$. The massless quark dispersion relations, for
the limit $k << M$, are
\be \omega(k)=M+\frac{k}{3}+\frac{k^2}{3M}+{\cal O}(k^3), \ee
\be \omega(k)=M-\frac{k}{3}+\frac{k^2}{3M}+{\cal O}(k^3), \ee
where $M$ is given by (5.~10). For the limit $k>>M$, the
dispersion relations are written as (4.~29) and (4.~30), with
$M_{L,R}^2=M^2$. The inclusion of non-leading contributions leads
to increase $M_{Lead}$ by $2.223\%$. We have checked the
validity of expressions (5.~11) and (5.~12) by comparing the
results obtained from them with the results obtained by means of
the exact numerical solution of the dispersion relations. The
relative error of the results obtained from (5.~11) and (5.~12)
is smaller than $0.01\%$ with respect to the exact numerical
results. Given the fit (5.~10), the solution of (5.~8) can be
well approximated by
\be \omega=M_{Lead}\left[1-2A^2\left(\gamma_{E}+\log\left[
\frac{M_{Lead}}{2T}\right]\right)\right]^{-1/2}. \ee
Expanding the root square in (5.~13), we can see
that the non-leading terms of the quark self-energy modify
$M_{Lead}$ as $\propto g_{s}^{(n)}T$, with $(n=3,5,...)$, in
accordance with the conclusions given in \cite{kobref}.

The Lorentz-invariant functions in the Landau gauge are
\be a^{\pounds}(\omega,k)=\frac{4}{3}g_{s}^2[A(0,0) +
A_0^{\delta}(0,0)],\ee
\be b^{\pounds}(\omega,k)=\frac{4}{3}g_{s}^2[B(0,0) +
B_0^{\delta}(0,0)],\ee
where $A(0,0)$ and $B(0,0)$ are given by (3.~21) and (3.~22)
with $M_1=M_2=0$. The integrals $A_0^{\delta}(0,0)$ and
$B_0^{\delta}(0,0)$ are given by (B.~9) and (B.~10) with $M_1=0$.
Substituting (5.~14) and (5.~15) into (2.~16), we compute the
quark dispersion relations in the Landau gauge. These dispersion
relations coincide with those at leading order ${\cal O}(g_{s}T)$
in the Feynman gauge, as is shown in Fig.~9. In other words, the
Lorentz-invariant functions (5.~14) and (5.~15) reduce to the
analytical form of (4.~21) and (4.~22) with $L_{I,F}=4g_{s}^2/3$.
This means that non-leading contributions of $A(0,0)$ and $B(0,0)$
in (5.~14) and (5.~15) are canceled exactly by $A_0^{\delta}(0,0)$
and $B_0^{\delta}(0,0)$ contributions, respectively. For
$k<<\omega$, we have analytically proved this fact. For the Landau
gauge, the gauge dependent parameter $\rho$ in (5.~10) is zero.
This result indicates that the one-loop dispersion relations
are only gauge invariant at leading order ${\cal O}(g_{s}T)$
\cite{welref, tayref}. As the thermal imaginary part of the
one-loop self-energy contains only non-leading terms, it is highly
gauge dependent. For a consistent calculation of this part, it
is necessary to go to the imaginary time formalism of the thermal
field theory and to use Pisarski's method for resumming thermal
loops \cite{pis1ref, schref}.

We will now study the massive quark case in QCD. The
Lorentz-invariant functions in the Feynman gauge (3.~18)-(3.~20)
contain terms dependent of $M_q$ . For the case $M_q<<\omega$,
it is possible to neglect\footnote{We have numerically proved
that neglecting the quark mass in the logarithmic integrals,
for $M_q<\omega/20$, leads to relative errors in the integrals
smaller than $0.1\%$.} $M_q$ in the logarithmic integrals,
following the same arguments as given in \cite{petref}.

We can write (3.~18)-(3.~20) for $M_q<<\omega$ as
\be a(\omega,k)=a^{Q}(\omega,k)+ A^2\frac{M_{q}^2}{k^2}\left[
2\left(\log\left[\frac{M_q}{\pi T}\right]+
\gamma_E \right)-\frac{1}{k}I_2\right], \ee
\be b(\omega,k)=b^{Q}(\omega,k)+ A^2\frac{M_{q}^2}{k^2}\left[
2\left(\log\left[\frac{M_q}{\pi T}\right]+
\gamma_E \right)-\frac{1}{k}I_2\right], \ee
\be c(M_{q},0)=-4A^2\frac{1}{k}I_2, \ee
where $a^{Q}(\omega,k)$ and $b^{Q}(\omega,k)$ are the
Lorentz-invariant functions for the massless case (5.~1) and
(5.~2), and $I_2$ is given by (5.~4). Substituting (5.~16)-(5.~18)
in (2.~15), taking the limits $k<<\omega$, $M_{q} <<\omega$, and
setting $k=0$, we obtain the following relation:
\be \omega^2\left[1-A^2 F_2(\omega,T)\right]-M_{Lead}^2=
\omega M_{q}\left[1-4A^2 F_2(\omega,T)\right], \ee
where $F_2(\omega,T)$ is given by (5.~9). This relation for
$M_{q}=0$ reduces to (5.~8). If non-leading contributions
manifested by means of $F_2(\omega,T)$ are set to zero, the
relation (5.~19) leads to the leading thermal effective masses
$M_{\pm}$ defined in (4.~13) and (4.~14).

Taking into account the validity of (5.~19) only for
$M_{q}<<M_{Lead}$, we obtain through a fit of the numerical
solutions of (5.~19), that
\be M_{\pm}(M_q)=M_{Lead}\left(1+\rho\left[1-\frac{M_q}{2T}\right]
\right) \pm \frac{M_q}{2}\left(1-\sigma\left[1 \mp \frac{M_q}{2T}
\right]\right),\ee
where $\rho=0.02223$ and $\sigma=0.135$. For $M_q=0$, (5.~20)
reduces to (5.~10). The dispersion relations given by (4.~13) and
(4.~14), with $M_{\pm}$ defined by (5.~20), contain all $T$
contributions from the one-loop self-energy. We have also checked
the validity of these analytical dispersion relations by comparing
the results obtained from them with those obtained sol-  ving
numerically the dispersion relations. For $M_q<M_{Lead}/20$, the
relative error is smaller than $0.01\%$.

The Lorentz-invariant functions in the Landau gauge for massive
quarks in QCD are given by
\be a^{\pounds}(\omega,k)=\frac{4}{3}g_{s}^2[A(M_{q},0) +
 A_0^{\delta}(M_{q},0)],\ee
\be b^{\pounds}(\omega,k)=\frac{4}{3}g_{s}^2[B(M_{q},0) +
B_0^{\delta}(M_{q},0)],\ee
\be c^{\pounds}(\omega,k)=\frac{5}{2}
\frac{4}{3}g_{s}^2C(M_{q},0),\ee
where $A(M_{q},0)$, $B(M_{q},0)$ and $C(M_{q},0)$ are defined by
(3.~21)-(3.~23), with $M_2=0$. $A_0^{\delta}(M_{q},0)$ and
$B_0^{\delta}(M_{q},0)$ are defined by (B.~9) and (B.~10),
respectively. The quark dispersion relations in the Landau gauge
are computed inserting (5.~21)-(5.~23) into (2.~15). These
dispersion relations coincide with those obtained at leading
order in the Feynman gauge, as is shown in Fig.~10. The gauge
dependent parameters $\rho$ and $\sigma$ in (5.~20) are zero in
the Landau gauge computation; consequently, the massive quark
dispersion relations are gauge invariant at leading order in
temperature ${\cal O}(g_{s}T)$ and in mass ${\cal O}(M_{q})$.

The inclusion of the non-leading one-loop contributions changes
the leading thermal effective masses ~according to the quark mass.
{}~For instance, the change in the leading thermal effective masses
for the $c$ quark at $T=100$ GeV are: $\delta M_{+}=0.0202M_{+}$
and $\delta M_{-}=0.0241M_{-}$. For the $s$ quark, they are:
$\delta M_{+}=0.0220M_{+}$ and $\delta M_{-}=0.0224M_{-}$.
Thus, the effect of the sub-leading contributions to the dispersion
relations is nearly $2\%$ for the light quarks.  We have not
obtained a similar fit of the thermal effective masses as (5.~20)
for the heavy quark case. We have checked numerically for this
case that the sub-leading contributions to the thermal effective
mass are small in comparison with the leading contributions.

For the MSM dispersion relations in the unbroken phase, the
situation is quite similar to the QCD case for massless quarks.
For instance, the thermal effective masses including leading and
non-leading contributions in the Feynman gauge are given by
(see Eq. (5.~10)):
\be M_{L,R}^{u}=M_{L,R}^{u(Lead)}(1 + \rho ), \ee
being $M_{L,R}^{u(Lead)}$ the leading contribution to the
thermal effective masses in the unbroken phase
\be M_{L,R}^{u(Lead)}=\sqrt{{\bar f}_{L,R}^{u} \frac{T^2}{8}},\ee
with ${\bar f}_{L,R}^{u}=f+f_{L,R}^{Z}+f_{H}+f_{L,R}^{W}$, where
the coefficients $f_{n}$ are given by Eqs. (4.~4)-(4.~7). The
Lorentz-invariant functions in the Landau gauge for this case are
presented in the Appendix C. The dispersion relations in the
Landau gauge coming from substituting (C.~1) and (C.~2) in (2.~13)
and (2.14). In a similar way as the QCD case, the branches in the
Landau gauge are superimposed on the Feynman gauge branches at
leading order. In consequence, the fermion dispersion relations
in the unbroken phase are gauge invariant at leading order. The
consideration of all terms in the quark self-energy introduces
a small gauge dependence in the dispersion relations. Taking
into account (5.~24), the sub-leading contribution to the thermal
effective mass in the Feynman gauge is
\be \delta M_{L,R}^{u} = \rho M_{L,R}^{u(Lead)},\ee
where $\rho \simeq 0.023$ for this case. This contribution
induces a gauge dependence in our nu\-me\-ri\-cal computations of
nearly $2\%$. By the mentioned general considerations this
sub-leading contribution is of order $\sim g_{s}^3 T$ in a
general gauge.

The results obtained for massive quarks in QCD can be extended
to the MSM in the broken case. The quark dispersion relations
are gauge invariant at leading order in tem\-pe\-ra\-tu\-re and
in quark mass. The non-leading order terms in the quark self-energy
introduce a small gauge dependence in the dispersion relations. We
have checked numerically that the thermal effective masses in the
broken phase $M^{b}_{L,R}$ can be written as:
\be M^{b}_{L,R}=\frac{M_{+}^{b}(M_q)_{L,R}+
M_{-}^{b}(M_q)_{L,R}}{2},\ee
where taking into account (5. 20), $M_{\pm}^{b}(M_q)_{L,R}$ are given
by:
\be M_{\pm}^{b}(M_q)_{L,R}=M_{L,R}^{b(Lead)}\left(1+\rho\left[1-
\frac{M_q}{2T}\right]\right) \pm \frac{M_q}{2}\left(1-\sigma\left
[1 \mp \frac{M_q}{2T}\right]\right),\ee
with $\sigma \simeq 0.14$, and $M_{L,R}^{b(Lead)}$ are the thermal
effective masses at leading order in the broken phase
\be M_{L,R}^{b(Lead)}=\sqrt{{\bar f}_{L,R}^{b}\frac{T^2}{8}},\ee
where the coefficients are given by
\be {\bar f}_{L,R}^{b}=f+(1-x_Z)f_{L,R}^{Z}+(1-x_H)f_{H}
+(1-x_W)f_{L,R}^{W}.\ee
In this equation $x_Z$, $x_H$, $x_W$ denote the effect of the
boson mass contributions on the thermal effective masses, being
$0< x_Z$, $x_H$, $x_W << 1$. This fact permits to understand why
the thermal effective masses in the broken phase $M^{b}_{L,R}$ are
smaller than those in the unbroken phase $M^{u}_{L,R}$. The
sub-leading contribution to the thermal effective masses in the
Feynman gauge, taking into account (5.~27), is
\be \delta M_{L,R}^{b}=\rho M_{L,R}^{b(Lead)}\left[1-\frac{M_q}{2T}
\right] + \sigma\frac{M_{q}^2}{4T},\ee
inducing a gauge dependence in our numerical computations about $2\%$
for the light quarks. Due the mentioned general grounds, the
sub-leading contribution to the thermal effective masses is of
order $\sim g_{s}^3[T - M_{q}]$ for quarks, and of order
$\sim g^3[T - M_{e}]$ for charged leptons and neutrinos in a general
gauge.

It is easy to prove that the gauge dependence induced in the
shift of the thermal effective masses between the two phases is
smaller than the one mentioned above. The difference between the
gauge dependent contributions to the thermal effective masses in
a general gauge is of order $\sim g_{s}^3 M_{q}$ for quarks and
of order $\sim g^3 M_{e}$ for charged leptons and neutrinos.
For the light quark case, the difference in the Feynman gauge is:
\be \delta M_{L,R}^{b}-\delta M_{L,R}^{u} \simeq -\rho
M_{L,R}^{b(Lead)} \frac{M_q}{2T}+ \sigma\frac{M_{q}^2}{4T}.\ee
This gauge dependence is smaller than (5. 26) and (5. 31).

\seCtion{Numerical results}

\hspace{2.0mm}
We now give the numerical results we have obtained for the
dispersion relations for up quarks $(u,c,t)$, down quarks $(d,s,b)$,
charged leptons $(e, \mu, \tau)$ and neutrinos
$(\nu_e, \nu_{\mu}, \nu_{\tau})$. For $T=100$ GeV, the following
values for the masses in GeV have been used: $M_W=52$, $M_Z=59$,
$M_H=100$, $M_d=0.006$, $M_s=0.09$, $M_b=2.9$, $M_u=0.003$,
$M_c=1.0$, $M_t=114.7$, $M_e=0.0003$, $M_{\mu}=0.069$,
$M_{\tau}=1.176$ and massless neutrinos. As we have mentioned, we
are specially interested in the low-momentum regime where the
collective behaviour is present. Adequate boson and fermion mass
values for each temperature have been used, for the cases where
the thermal dependence has been studied. The values of the
coupling constants have been fixed to $\alpha_s=0.1$,
$\alpha_w=0.035$ for all temperatures.

\vspace{2.0mm}
In Figs.~3 and 4, we show the branches for quarks $s$ and
$c$ respectively in both phases. In the unbroken phase, the
left-handed branches are above the right-handed ones. Although
the branches in the broken phase are not associated to a specific
chirality, the number of branches is also four. While in the
unbroken phase the left-handed abnormal branch is crossed by the
right-handed normal branch, in the broken phase there is not
crossing between the branches. In the broken phase, the branches
that now have a hyperbolic form are separated between the minimum
and the maximum by an energy gap of order $\sim M_f$. The
quasi-particle excitations cannot exist in this energy range.

\vspace{2.0mm}
The branches for each quark flavour are different in the broken
phase and clearly the fermionic spectrum described by them is
flavour non-degenerate. In the unbroken phase, the left-handed
branches for the different quark flavours are numerically the
same in our computation, but a very small difference should exist
for them. The fermionic spectrum is flavour non-degenerate,
but the differences for different flavours are very small. The
fermionic spectrum described by the right-handed branches is also
flavour non-degenerate in a similar way to the left-handed
fermionic spectrum.

\vspace{2.0mm}
Since temperature and mass corrections are considered, the thermal
effective masses in the broken phase are smaller than those in the
unbroken phase. For the $s$ quark case, for instance (see Fig.~3),
the thermal effective mass of the left-handed branches in the
unbroken phase ~is $\simeq 51.4$ GeV, ~while that for ~the
corresponding branches in the ~broken phase is $\simeq 50.0$ GeV.
The shift of the thermal effective masses for the lower branches is
also present, but is smaller. In our perturbative calculation, this
effect can be understood through the presence of the massive boson
propagator (3.~2) in the fermion self-energy. This discontinuity of
the thermal effective masses between the two sides of the boundary
created during the first order electroweak phase transition has
recently been noted in \cite{monref}. In this reference, using a
numerical simulation of the phase transition in the SU(2) Higgs model
on the lattice, was found that the thermal effective mass for the
$W$ and Higgs boson is smaller in the broken phase than in the
unbroken one.

\vspace{2.0mm}
The dependence with $T$ of the top quark thermal effective masses
for the branches in the broken phase is presented in Fig.~5.
Two branches go down appreciably as the temperature is lowered
($M_t$ increases). For $T$ below $120$ GeV, these branches can
become unphysical as is suggested by this figure.\footnote{Because
of numerical problems, it was not possible to obtain results for
the branches for $T<155$ GeV.} The special behaviour of the top
quark branches is obtained because this quark acquire a large mass
during the electroweak phase transition.

\vspace{2.0mm}
The branches for the $\tau$ lepton are shown in Fig.~6. An energy
gap between the hyperbolic branches in the broken phase is also
present. As in the quark case, the thermal effective masses for
charged leptons in the broken phase are smaller than those in the
unbroken one. In the broken phase, the branches for each flavour
are different and consequently the fermionic spectrum is flavour
non-degenerate. In the unbroken phase, the branches of a specific
chirality are the same for different flavours within the precision
of the present computation. The fermionic spectrum described is
flavour non-degenerate, but the differences for different flavours
are very small.

\vspace{2.0mm}
The branches for the electron neutrino are shown in Fig.~7. There
are only two branches because the left-handed chirality of the
neutrino. In both phases, the branches for different flavours are
the same in our computations. However, the fermionic spectrum is
flavour non-degenerate, in both phases, but the differences are
very small for different flavours. The thermal effective masses
for neutrinos in the broken phase are smaller than those in the
unbroken one. The $T$ dependence of the electron neutrino thermal
effective mass is presented in Fig.~8. The thermal effective mass
is zero for $T=0$, as can be concluded from this figure.

\seCtion{Conclusions}

\hspace{3.0mm}
One-loop dispersion relations for the different fermionic sectors
of the MSM were obtained. The calculation was performed in the
real time formalism of the thermal field theory, in the Feynman
gauge. Numerical results were presented for temperatures of the
plasma near the critical one of the electroweak phase transition.
For all the fermions in the broken phase and in the unbroken one,
the fermionic spectrum described by the branches of the dispersion
relations is flavour non-degenerate.

\vspace{2.0mm}
In the unbroken phase, the fermionic quasi-particles can be
associated to a specific chi\-ra\-li\-ty because the chiral symmetry
is preserved by the MSM before the ESSB. Since the chiral symmetry
is violated by the MSM after the ESSB, the quasi-particles cannot
be associated to a specific chirality in the broken phase for the
case of massive fermions. The number of branches for quarks and
charged leptons is four in both phases, while it is only two for
neutrinos.

\vspace{2.0mm}
The thermal effective masses for all fermions in the broken phase
are smaller than those in the unbroken one. The existence of this
shift for the thermal effective masses can be understood through
the presence of the massive boson propagators in the fermion
self-energy. A similar shift for the $W$ and Higgs boson thermal
effective mass has recently been observed in \cite{monref}, but
through a non-perturbative lattice calculation. For quarks and
charged leptons in the broken phase, an energy gap of order
$\sim M_f$ was obtained between the branches that have a hyperbolic
form. This gap appears as a consequence of the electroweak
contributions to the dispersion relations.

\vspace{2.0mm}
The $T$ dependence of the thermal effective masses for the top
quark and for the electron neutrino was studied numerically. The
results for the massive top quark suggest that two branches
become unphysical in the broken phase for $T<120$ GeV.

\vspace{2.0mm}
We have studied the gauge dependence of the dispersion relations
for the massless and massive quark cases in QCD by confronting
computations in the Feynman and the Landau gauges. We have found
for both cases that the dispersion relations in the Landau gauge
are equal to those obtained at leading order in the Feynman gauge.

\vspace{2.0mm}
The fermion dispersion relations in the MSM include leading and
sub-leading contributions. For the unbroken phase, in a similar way
as the massless QCD case, the branches in the Landau gauge are
superimposed on the Feynman gauge branches at leading order.
Consequently, the fermion dispersion relations are gauge invariant
at leading order. The consideration of the non-leading terms in
the fermion self-energy introduces a small sub-leading gauge
dependence in the dispersion relations. General considerations
indicate that the sub-leading contribution to the thermal effective
masses, in a general gauge, is of order $\sim g_{s}^3 T$ for
quarks and $\sim g^3 T$ for charged leptons and neutrinos.
Particularly, the sub-leading contribution in the Feynman gauge is
$\delta M_{L,R}^{u}=\rho M_{L,R}^{u(Lead)}$ for the quark case, where
$M_{L,R}^{u(Lead)}$ is the thermal effective mass at leading order
in the unbroken phase (see Eqs. (5.~25)) and $\rho \simeq 0.023$.
The sub-leading contribution induces a gauge dependence in our
numerical computations of nearly $2\%$.

\vspace{2.0mm}
The fermion dispersion relations in the broken phase are gauge
invariant at leading order in temperature and in fermion mass. The
small sub-leading contribution to the dispersion relation is
gauge dependent. On general grounds, the sub-leading contribution
to the thermal effective masses, in a general gauge, is of order
$\sim g_{s}^3[T - M_{q}]$ for quarks, and of order
$\sim g^3[T - M_{e}]$ for charged leptons and neutrinos. The
sub-leading contribution in the Feynman gauge is
$ \delta M_{L,R}^{b}(M_q)=\rho M_{L,R}^{b(Lead)}\left[1-
\frac{M_q}{2T}\right] + \sigma\frac{M_{q}^2}{4T}$, where
$M_{L,R}^{b(Lead)}$ are the thermal effective masses at leading
order in the broken phase (see Eq. (5. 29)), and the gauge parameters
$\rho \simeq 0.023$ and $\sigma \simeq 0.14$. The gauge dependence
induced in our numerical computations is about $2\%$ for the light
quark case.

\vspace{2.0mm}
The gauge dependence induced in the shift of the thermal effective
masses between the two phases is smaller than the one mentioned
above. The difference between the gauge dependent contributions to
the thermal effective masses in a general gauge is of order
$\sim g_{s}^3 M_{q}$ for quarks and of order $\sim g^3 M_{e}$ for
charged leptons and neutrinos. This difference for light quarks in
the Feynman gauge is:
$ \delta M_{L,R}^{u}-\delta M_{L,R}^{b} \simeq \rho
M_{L,R}^{b(Lead)} \frac{M_q}{2T}- \sigma\frac{M_{q}^2}{4T}$. In
our numerical computations is true that
$\delta M_{L}^{u}-\delta M_{L}^{b} << M_{L}^{u}-M_{L}^{b}$,
showing that the gauge dependence in the shift is very small.

\section*{Acknowledgements}

\hspace{3.0mm}
We are indebted to Bel\'en Gavela and Olivier P\`ene for important
suggestions and comments about the elaboration of the present work.
We also thank Alvaro de R\'ujula, Pilar Hern\'andez, Jean Orloff,
Juan Carlos Pinto, Nuria Rius and Miguel Angel V\'azquez-Mozo for
many helpful conversations. C.~Quimbay would like to thank
COLCIENCIAS (Colombia) for financial support. S.~Vargas-Castrill\'on
would like to thank Consejer\'{\i}a de Educaci\'on y Cultura de la
Comunidad Aut\'onoma de Madrid for financial support.

\section*{Appendices}
\appendix
\seCtion{Calculation of $Z$-tadpole diagram}

\hspace{3.0mm}
The generic $Z$-tadpole diagram in Fig.~1 is given by:
\be \Sigma_{tad}(K)= -i\frac{g^2}{C_W^2}\gamma_{\mu}
\Pi_E D_Z^{\mu \nu}(0) \int\frac{d^{4}p}
{(2\pi)^4}Tr[\gamma_{\nu}\Pi_I S(p)],\ee
where $S(p)$ is the fermionic finite temperature propagator (2.~1),
$D_Z^{\mu \nu}(0)$ is the propagator of the $Z$ boson at momentum
transfer equal to zero:
\be D_Z^{\mu \nu}(0)=\frac{g^{\mu \nu}}{M_Z^2},\ee
and $\Pi_X=(T_X^3 - Q_X S_W^2)L - Q_X S_W^2 R$, are the chiral
operators of the tadpole vertices, with $X=E,I$. Here, $E$ and
$I$ refers to external and internal fermion respectively. The
expression (A.~1) has the same form as that shown by N\"{o}tzold
and Raffel in \cite{rafref} for the particular case of neutrino
propagation.

Taking over the expressions of the propagators (2.~1) and (A.~2)
in (A.~1) and only considering the real part contribution
$\mbox{Re}\,\Sigma'_{tad}(K)$ of the finite temperature
contribution $\Sigma'_{tad}(K)$, we obtain:
\be \mbox{Re}\,\Sigma'_{tad}(K)= \frac{g^2}{C_W^2}\frac
{\Pi_E \Omega_I}{M_Z^2}\int\frac{d^{4} p}{(2\pi)^3}p
{\hspace{-1.9mm}\slash}\delta(p^2-M_i^2)n_f(p),\ee
where $n(f)$ is the Fermi-Dirac distribution function given by
(2.~6) and $\Omega_I=(T_I^3-Q_I S_W^2)/2$. We follow the same
procedure as the one used for the evaluation of the generic
diagrams of Fig.~2. If we multiply (A.~3) by either
$K{\hspace{-3.1mm}\slash}$ or $u{\hspace{-2.2mm}\slash}$, and then
take the trace of the product, we obtain:
\be \frac{1}{4}Tr(K{\hspace{-3.1mm}\slash}\,\mbox{Re}\,
\Sigma'_{tad})=\frac{g^2}{C_W^2}\frac{\Pi_E \Omega_I}{M_Z^2}\
\frac{dp_0 d^{3} p}{(2\pi)^3}(\omega p_0 - \vec k \cdot \vec p)
\delta(p_0^2-|\vec p|^2-M_I^2)n_f(p),\ee
\be \frac{1}{4}Tr(u{\hspace{-2.2mm}\slash}\,\mbox{Re}\,
\Sigma'_{tad})=\frac{g^2}{C_W^2}\frac{\Omega_i \Pi_I}{M_Z^2}\int
\frac{dp_0 d^{3} p}{(2\pi)^3} p_0
\delta(p_0^2-|\vec p|^2-M_I^2)n_f(p).\ee

The argument of the delta-Dirac functions of the expressions
(A.~4) and (A.~5) have the zeros when
$p_0=\pm (|\vec p|^2 + M_I^2)^{1/2}=\epsilon_I$.
When we integrate over $p_0$, we obtain the positive and the
negative contributions of $\epsilon_I$. Immediately, one can see
that (A.~5) is zero and the integral (A.~4) is seen to be zero
after the angular integration. It is direct to prove that if we
take the trace of (A.~3), the result vanishes. As a consequence,
the contribution of the $Z$-tadpole diagram to the real part of
the fermionic thermal self-energy is zero. This result differs
from that given in \cite{nieref}, where the tadpole diagram
contributes to the real part of the neutrino thermal self-energy.
This contribution exists because the Fermi-Dirac distribution
function used there was defined without the absolute value that
we used in (3.~6).

\seCtion{Real part of the fermion self-energy}
\begin{Large}
\hspace{1.2cm}{\bf in the Landau gauge}
\end{Large}

\hspace{3.0mm}

The thermal propagator in the Landau gauge for massless gauge
$D_{\mu \nu}^{\pounds}(p)$ and massless scalar $D^{\pounds}(p)$
bosons are:
\be D_{\mu \nu}^{\pounds}(p)=\left[-g_{\mu\nu}+
\frac{p_{\mu}p_{\nu}}{p^2}\right]\left[\frac{1}
{p^2+i\epsilon}-i{\Gamma}_b(p)\right],\ee
\be D^{\pounds}(p)=\frac{1}{p^2+i\epsilon}-i{\Gamma}_b(p),\ee
with ${\Gamma}_b(p)$ defined by (3.~5). The contribution to the
self-energy from the generic gauge boson one-loop diagram of
Fig.~2a is:
\be \Sigma^{\pounds}(K)= ig^2 C_c \int\frac{d^{4} p}{(2\pi)^4}
D_{\mu \nu}^{\pounds}(p) {\gamma}^{\mu}S(p+K){\gamma}^{\nu},\ee
where $S(p)$ is given by (3.~1). Inserting the expressions
(3.~1) and (B.~1) into (B.~3) and keeping only the real part
$\mbox{Re}\,\Sigma'^{\pounds}(K)$ of the finite temperature
contribution $\Sigma'^{\pounds}(K)$ to the self-energy, it is
possible to write:
\be \mbox{Re}\,\Sigma'^{\pounds}(K)= \mbox{Re}\,\Sigma'(K) +
\mbox{Re}\,\delta\Sigma'(K),\ee
where $\mbox{Re}\,\Sigma'(K)$ is the real part of the finite
temperature contribution to the self-energy in the Feynman gauge
(3.~8) and $\mbox{Re}\,\delta\Sigma'(K)$ is given by:
\vspace{3.0mm}
\be \mbox{Re}\,\delta\Sigma'_0= g^2C_c \int
\frac{d^{4}p}{(2\pi)^4}p{\hspace{-1.9mm}\slash}(p{\hspace{-1.9mm}
\slash}+K{\hspace{-3.1mm}\slash}+M_1)p{\hspace{-1.9mm}\slash}
\left[\frac{{\Gamma}_b (p)}{p^2[(p+K)^2-M_1^2]}-
\frac{{\Gamma}_f(p+K)}{p^4}\right],\ee
where the denominators are defined by their principal value.
We note that there is a principal-value singularity $1/p^2$ in
the longitudinal boson propagator. Its product with $\delta(p^2)$
is defined by $-\delta(p^2)/p^2 \rightarrow \delta'(p^2)$, where
the prime denotes the derivative with respect to $p_0^2$
\cite{welref}. Using this definition and following a procedure
similar to the preceding case, the traces
$\frac{1}{4}Tr(K{\hspace{-3.1mm}\slash}\,\mbox{Re}\,
\delta\Sigma'_0)$ and
$\frac{1}{4}Tr(u{\hspace{-2.2mm}\slash}\,\mbox{Re}\,
\delta\Sigma'_0)$ are calculated.
With similar expressions to (3.~15)-(3.~17), the following
relations are obtained:
\vspace{3.0mm}
\be a_0^{\delta}(\omega,k)=g^2C_c A_0^{\delta}(M_1,0),\ee
\be b_0^{\delta}(\omega,k)=g^2C_c B_0^{\delta}(M_1,0),\ee
\be c_0^{\delta}(\omega,k)=\frac{g^2C_c}{2} \frac{M_1}{M_I}
C(M_1,0),\ee
where:
\bea A_0^{\delta}(M_1,0)=\int^\infty_0\frac{dp}{4\pi^2}
\frac{1}{k^2}\left(\left[\frac{\omega^2-k^2-M_1^2}{2p}
+\frac{\omega^2-k^2-M_1^2}{2T}e^{P/T}n_B(p)\left(1+2
\frac{\omega}{k}L_2^{-}(p)\right)\right.\right.\nn\\
-\frac{1}{2}\left(\omega R^{-}(p)+pR^{+}(p)\right)
-\omega p\left(\omega T^{+}(p)+pT^{-}(p)\right)\nn\\
-\left.\left(\frac{\omega^2-k^2-M_1^2}{8k}
-\frac{(\omega^2-k^2-M_1^2)^2}{16kp}\left[\frac{1}{p}
+\frac{1}{T}e^{p/T}n_B(p)\right]\right)L_2^{+}(p)\right]
n_B(p)\nn\\
+\left.\frac{p}{\epsilon_1}\left[\frac{\omega^2+k^2}
{8k}L_1^{+}(p)-\frac{1}{2}pS^{+}(p)+\omega \epsilon_1 pU^{-}(p)+
\omega^2pV^{+}(p)\right]n_{F}(\epsilon_1)\right).\quad\quad\eea

\bea B_0^{\delta}(M_1,0)=-\int^\infty_0\frac{dp}{4\pi^2}
\frac{1}{k^2}\left(\left[\frac{\omega (\omega^2-k^2-M_1^2)}{2p}
\right.\right.\quad\quad\quad\quad\quad\quad\quad\quad\quad\quad
\quad\quad\quad\quad\quad\quad\nn\\
+\frac{\omega^2-k^2-M_1^2}{2T}e^{P/T}n_B(p)\left(\omega +2
\frac{(\omega^2 - k^2)}{k}L_2^{-}(p)\right)\nn\\
-\frac{1}{2}\omega \left(\omega R^{-}(p)+pR^{+}(p)\right)
-(\omega^2-k^2)p\left(\omega T^{+}(p)+pT^{-}(p)\right)\nn\\
-\left.\left(\frac{\omega(\omega^2-k^2-M_1^2)}{8k}
-\frac{\omega (\omega^2-k^2-M_1^2)^2}{16kp}\left[\frac{1}{p}
+\frac{1}{T}e^{p/T}n_B(p)\right]\right)L_2^{+}(p)\right]
n_B(p)\nn\\
+\left.\frac{p}{\epsilon_1}\left[\frac{\omega(\omega^2-k^2)}
{8k}L_1^{+}(p)+\frac{\omega}{2}pS^{+}(p)+\epsilon_1
(\omega^2 - k^2)pU^{-}(p)+\right]n_{F}(\epsilon_1)\right).
\quad\quad\eea
$C(M_1,0)$ is given by (3.~23) with $M_2=0$. The functions defined
in (B.~9) and (B.~10) are:
\be R^{\pm}(p)=\frac{(\omega^2-k^2-M_1^2)^2}
{(\omega^2-k^2-M_1^2 + 2\omega p)^2-4k^2p^2} \pm
\frac{(\omega^2-k^2-M_1^2)^2}
{(\omega^2-k^2-M_1^2 - 2\omega p)^2-4k^2p^2},\ee
\be S^{\pm}(p)=\frac{(\omega^2-k^2)(\omega^2-k^2+M_1^2)}
{(\omega^2-k^2+M_1^2 + 2\omega \epsilon_1)^2-4k^2p^2} \pm
\frac{(\omega^2-k^2)(\omega^2-k^2+M_1^2)}
{(\omega^2-k^2+M_1^2 - 2\omega \epsilon_1)^2-4k^2p^2},\ee
\be T^{\pm}(p)=\frac{(\omega^2-k^2-M_1^2)}
{(\omega^2-k^2+M_1^2 + 2\omega \epsilon_1)^2-4k^2p^2} \pm
\frac{(\omega^2-k^2-M_1^2)}
{(\omega^2-k^2+M_1^2 - 2\omega \epsilon_1)^2-4k^2p^2},\ee
\be U^{\pm}(p)=\frac{(\omega^2-k^2)}
{(\omega^2-k^2+M_1^2 + 2\omega \epsilon_1)^2-4k^2p^2} \pm
\frac{(\omega^2-k^2)}
{(\omega^2-k^2+M_1^2 - 2\omega \epsilon_1)^2-4k^2p^2},\ee
\be V^{\pm}(p)=\frac{(\omega^2-k^2+M_1^2)}
{(\omega^2-k^2+M_1^2 + 2\omega \epsilon_1)^2-4k^2p^2} \pm
\frac{(\omega^2-k^2+M_1^2)}
{(\omega^2-k^2+M_1^2 - 2\omega \epsilon_1)^2-4k^2p^2},\ee
and the functions $L_1^{\pm}$ and $L_2^{\pm}$ given by
(3.~13)-(3.~14) with $M_2=0$.

The Lorentz-invariant functions in the Landau gauge are given by
\be a^{\pounds}(\omega,k)=g^2C_c A^{\pounds}(M_1,0),\ee
\be b^{\pounds}(\omega,k)=g^2C_c B^{\pounds}(M_1,0),\ee
\be c^{\pounds}(\omega,k)=\frac{5}{2}g^2C_c \frac{M_1}{M_I}
C(M_1,0),\ee
where
\be A^{\pounds}(M_1,0)=A(M_1,0)+A_0^{\delta}(M_1,0),\ee
\be B^{\pounds}(M_1,0)=B(M_1,0)+B_0^{\delta}(M_1,0),\ee
with $A(M_1,0)$, $B(M_1,0)$ and $C(M_1,0)$ given by (3.~21)-(3.~23),
with $M_2=0$. $A_0^{\delta}(M_1,0)$ and $B^{\pounds}(M_1,0)$
are given by (B.~9) and (B.~10).

\seCtion{Lorentz-invariant functions in the Landau gauge}

\hspace{3.0mm}
The chiral projections of the Lorentz-invariant functions for the
quarks in the unbroken phase are given by:
\be a_{\stackrel{L}{R}}^{\pounds}(\omega,k)_{IF}=
\bar f_{\stackrel{L}{R}}^{1}A(0,0)+
\bar f_{\stackrel{L}{R}}^{2}A_0^{\delta}(0,0),\ee
\be b_{\stackrel{L}{R}}^{\pounds}(\omega,k)_{IF}=
\bar f_{\stackrel{L}{R}}^{1}B(0,0)+
\bar f_{\stackrel{L}{R}}^{2}B_0^{\delta}(0,0),\ee
where the integrals $A(0,0)$ and $B(0,0)$ are given by (3.~21) and
(3.~22), with $M_1=M_2=0$. The integrals $A_0^{\delta}(0,0)$ and
$B_0^{\delta}(0,0)$ are defined by (B.~9) and (B.~10), with $M_1=0$.
The coefficients are:
\be \bar f_L^{1}=8\left\{\left(\frac{2\pi\alpha_S}{3}
+\frac{3\pi\alpha_W}{8}\left[1+\frac{\tan^2\theta_W}{27}
+\frac{1}{3}\lambda_I^2\right]\right)\delta_{IF}
+\frac{\pi\alpha_W}{8} K_{Ii}^{+}\lambda_i^2 K_{iF}\right\},\ee
\be \bar f_L^{2}=8\left\{\frac{2\pi\alpha_S}{3}
+\frac{3\pi\alpha_W}{8}\left[1+\frac{\tan^2\theta_W}{27}\right]
\right\}\delta_{IF}, \ee
\be \bar f_R^{1}=8\left\{\frac{2\pi\alpha_S}{3}+\frac{\pi\alpha_W}{2}
\left[Q_I^2\tan^2 \theta_W+\frac{1}{2}\lambda_I^2\right]\right\}
\delta_{IF},\ee
\be \bar f_R^{2}=8\left\{\frac{2\pi\alpha_S}{3}+\frac{\pi\alpha_W}{2}
\left[Q_I^2\tan^2 \theta_W\right]\right\}\delta_{IF}.\ee

The chiral projections of the Lorentz-invariant functions for
charged leptons are:
\be a_{\stackrel{L}{R}}^{\pounds}(\omega,k)=
\bar d_{\stackrel{L}{R}}^{1}A(0,0)+
\bar d_{\stackrel{L}{R}}^{2}A_0^{\delta}(0,0),\ee
\be b_{\stackrel{L}{R}}^{\pounds}(\omega,k)=
\bar d_{\stackrel{L}{R}}^{1}B(0,0)+
\bar d_{\stackrel{L}{R}}^{2}B_0^{\delta}(0,0),\ee
where the integrals $A(0,0)$ and $B(0,0)$ are given by (3.~21) and
(3.~22), with $M_1=M_2=0$. The integrals $A_0^{\delta}(0,0)$ and
$B_0^{\delta}(0,0)$ are defined by (B.~9) and (B.~10), with $M_1=0$.
The coefficients are:
\be \bar d_L^{1}=\pi\alpha_W \left[ 2 + \frac{1}{c_w^2}
+\lambda_e^2\right],\ee
\be \bar d_L^{2}=\pi\alpha_W \left[ 2 + \frac{1}{c_w^2}\right],\ee
\be \bar d_R^{1}= 4\pi\alpha_W \left[\tan^2\theta_W +
\frac{\lambda_e^2}{2}\right],\ee
\be \bar d_R^{2}= 4\pi\alpha_W \left[\tan^2\theta_W \right].\ee

The chiral projections of the Lorentz-invariant functions for
neutrinos are:
\be a_L^{\pounds}(\omega,k)=\bar v^1 A(0,0)+
\bar v^2 A_0^{\delta}(0,0),\ee
\be b_L^{\pounds}(\omega,k)=\bar v^1 B(0,0)+
\bar v^2 B_0^{\delta}(0,0),\ee
where the integrals $A(0,0)$ and $B(0,0)$ are given by (3.~21) and
(3.22), with $M_1=M_2=0$. The integrals $A_0^{\delta}(0,0)$ and
$B_0^{\delta}(0,0)$ are defined by (B.~9) and (B.~10), with $M_1=0$.
The coefficients are:
\be \bar v^1 =\pi\alpha_W \left[ 2 + \frac{1}{c_w^2}
+\lambda_e^2\right],\ee
\be \bar v^2 =\pi\alpha_W \left[ 2 + \frac{1}{c_w^2}\right].\ee

\newpage

\begin{figure}[h]
\let\picnaturalsize=N
\def\picsize{3in}
\def\picfilename{fig1.ps}
\ifx\nopictures Y\else{\ifx\epsfloaded Y\else\input epsf \fi
\let\epsfloaded=Y
\centerline{\ifx\picnaturalsize N\epsfxsize \picsize\fi
\epsfbox{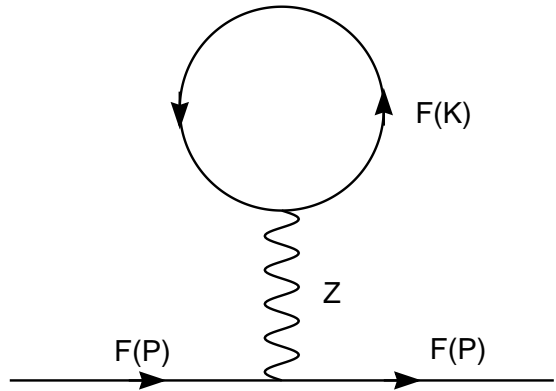}}}\fi
\caption{Generic $Z$-tadpole diagram}
\label{fig1}
\end{figure}

\newpage

\begin{figure}[h]
\let\picnaturalsize=N
\def\picsize{3in}
\def\picfilename{fig2.ps}
\ifx\nopictures Y\else{\ifx\epsfloaded Y\else\input epsf \fi
\let\epsfloaded=Y
\centerline{\ifx\picnaturalsize N\epsfxsize \picsize\fi
\epsfbox{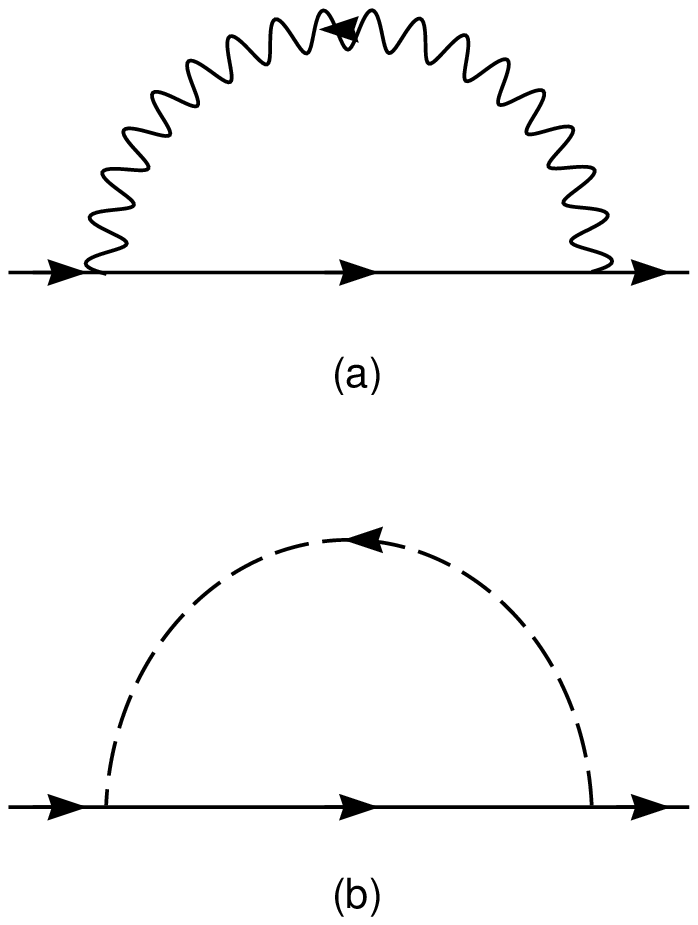}}}\fi
\caption{Generic gauge and scalar boson diagrams}
\label{fig2}
\end{figure}

\newpage

\begin{figure}[3]
\let\picnaturalsize=N
\def\picsize{5in}
\def\picfilename{fig3.ps}
\ifx\nopictures Y\else{\ifx\epsfloaded Y\else\input epsf \fi
\let\epsfloaded=Y
\centerline{\ifx\picnaturalsize N\epsfxsize \picsize\fi
\epsfbox{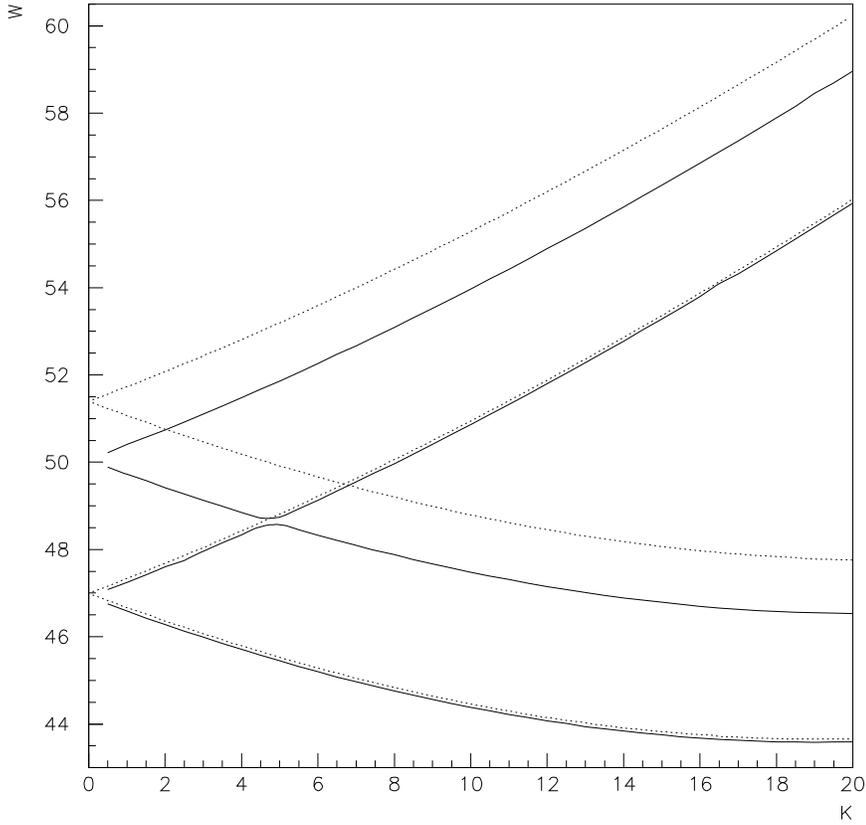}}}\fi
\caption{Dispersion relations at $T=100$ GeV for the $s$ quark in
the broken (solid curves) and the unbroken (dotted curves) phases,
neglecting quark mixing. The upper (lower) dotted curves correspond
to left (right) chirality branches. An energy gap ($\simeq 0.1$ GeV)
appears between the minimum and maximum of the hyperbolic branches.}
\label{fig3}
\end{figure}

\newpage

\begin{figure}[4]
\let\picnaturalsize=N
\def\picsize{5in}
\def\picfilename{fig4.ps}
\ifx\nopictures Y\else{\ifx\epsfloaded Y\else\input epsf \fi
\let\epsfloaded=Y
\centerline{\ifx\picnaturalsize N\epsfxsize \picsize\fi
\epsfbox{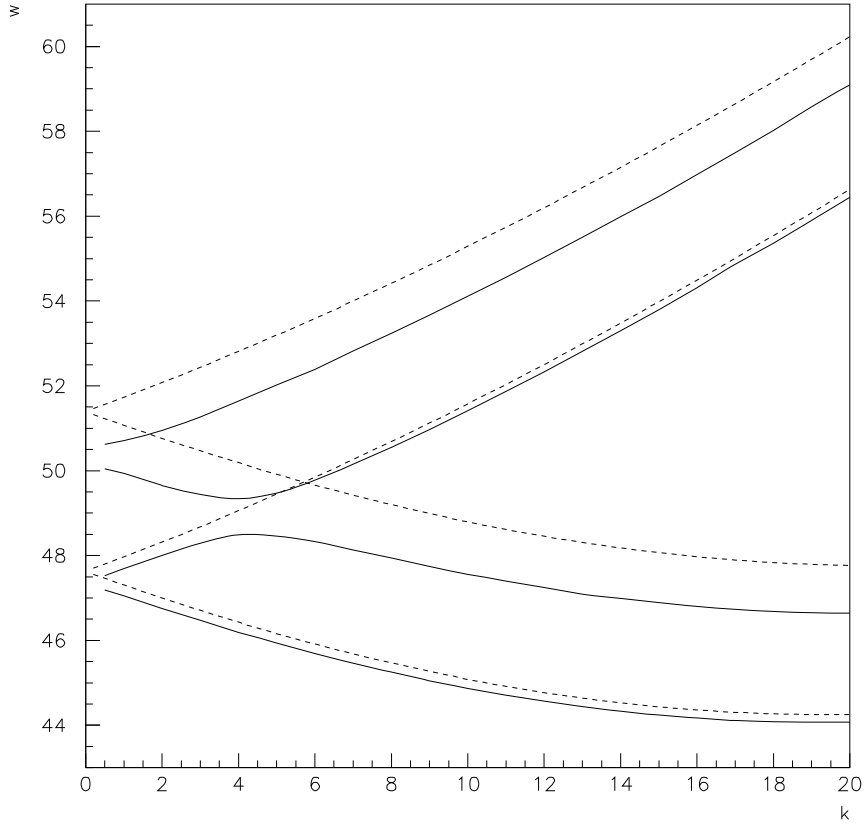}}}\fi
\caption{Dispersion relations at $T=100$ GeV for the $c$ quark in
the broken (solid curves) and the unbroken (dashed curves) phases,
neglecting quark mixing. The energy gap between the hyperbolic
branches is $\simeq 1.0$ GeV. The thermal effective masses for the
branches in the broken phase are smaller that those in the
unbroken phase.}
\label{fig4}
\end{figure}

\newpage

\begin{figure}[5]
\let\picnaturalsize=N
\def\picsize{5in}
\def\picfilename{fig5.ps}
\ifx\nopictures Y\else{\ifx\epsfloaded Y\else\input epsf \fi
\let\epsfloaded=Y
\centerline{\ifx\picnaturalsize N\epsfxsize \picsize\fi
\epsfbox{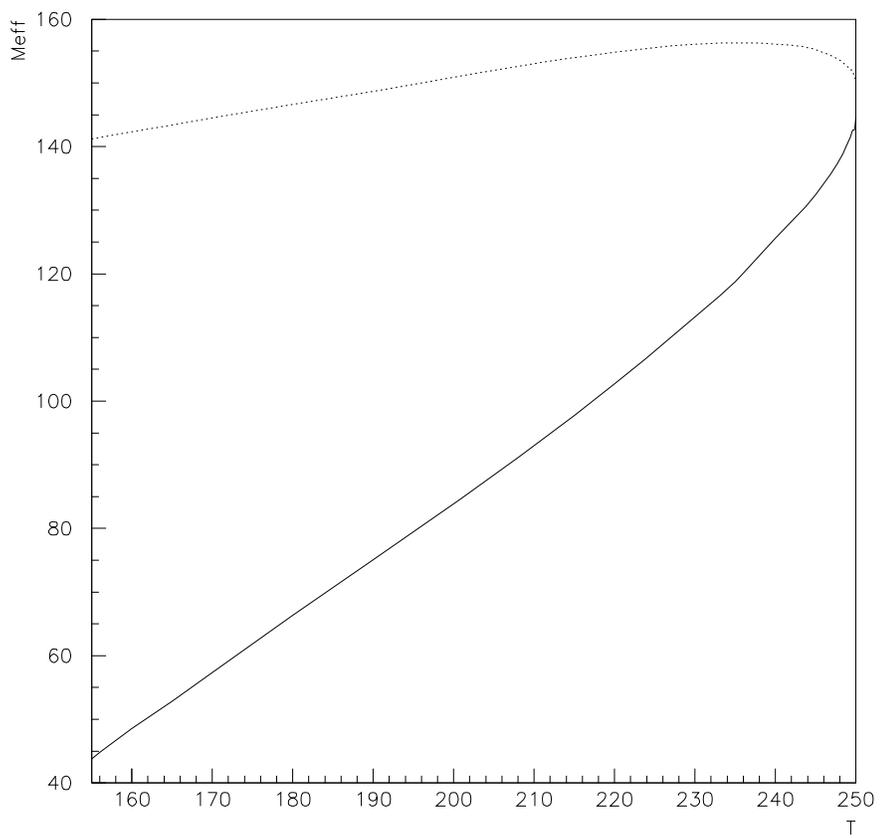}}}\fi
\caption{Variation with temperature of the thermal effective
masses for the top quark in the broken phase.}
\label{fig5}
\end{figure}

\newpage

\begin{figure}[6]
\let\picnaturalsize=N
\def\picsize{5in}
\def\picfilename{fig6.ps}
\ifx\nopictures Y\else{\ifx\epsfloaded Y\else\input epsf \fi
\let\epsfloaded=Y
\centerline{\ifx\picnaturalsize N\epsfxsize \picsize\fi
\epsfbox{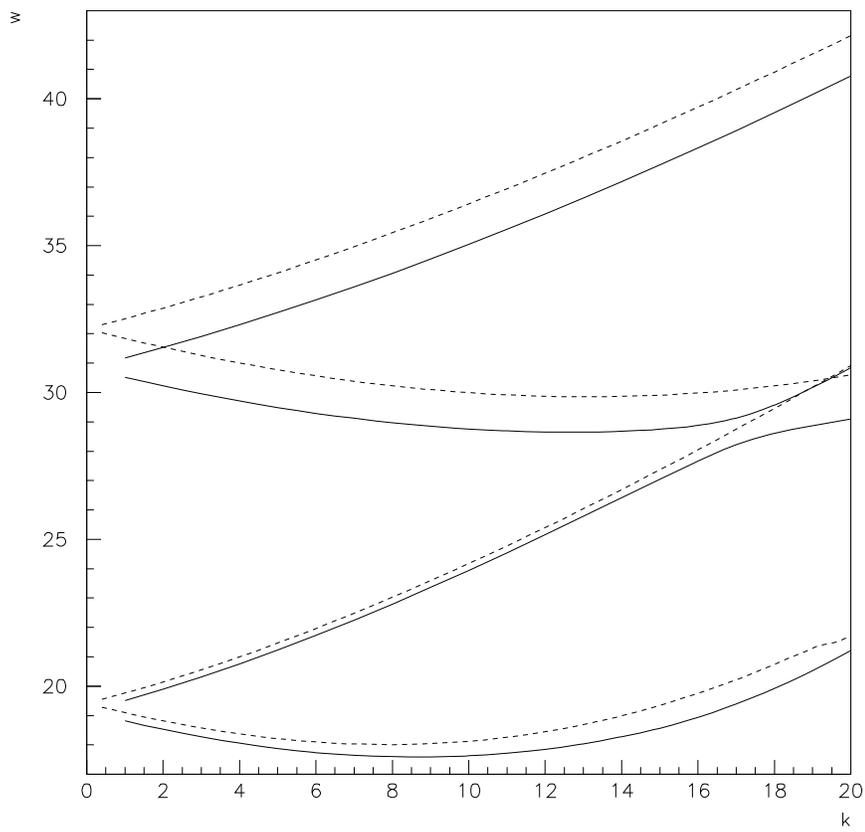}}}\fi
\caption{Dispersion relations at $T=150$ GeV for the $\tau$
lepton in the broken (solid curves) and the unbroken
(dashed curves).}
\label{fig6}
\end{figure}

\newpage

\begin{figure}[7]
\let\picnaturalsize=N
\def\picsize{5in}
\def\picfilename{fig7.ps}
\ifx\nopictures Y\else{\ifx\epsfloaded Y\else\input epsf \fi
\let\epsfloaded=Y
\centerline{\ifx\picnaturalsize N\epsfxsize \picsize\fi
\epsfbox{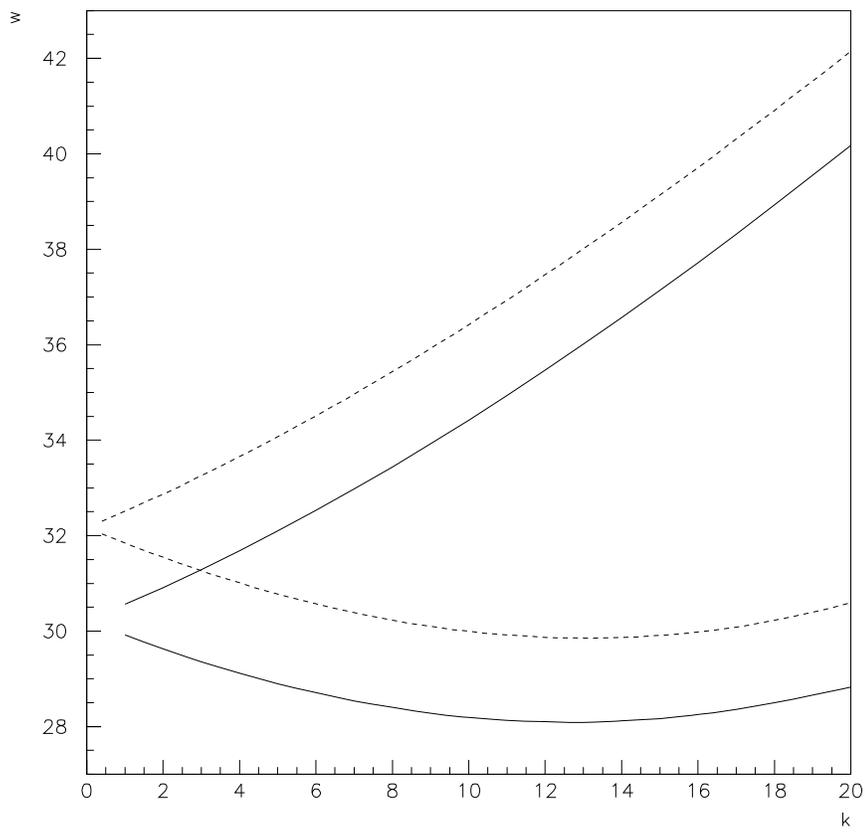}}}\fi
\caption{Dispersion relations at $T=150$ GeV for the electron
neutrino in the broken (solid curves) and the unbroken
(dashed curves) phases.}
\label{fig7}
\end{figure}

\newpage

\begin{figure}[8]
\let\picnaturalsize=N
\def\picsize{5in}
\def\picfilename{fig8.ps}
\ifx\nopictures Y\else{\ifx\epsfloaded Y\else\input epsf \fi
\let\epsfloaded=Y
\centerline{\ifx\picnaturalsize N\epsfxsize \picsize\fi
\epsfbox{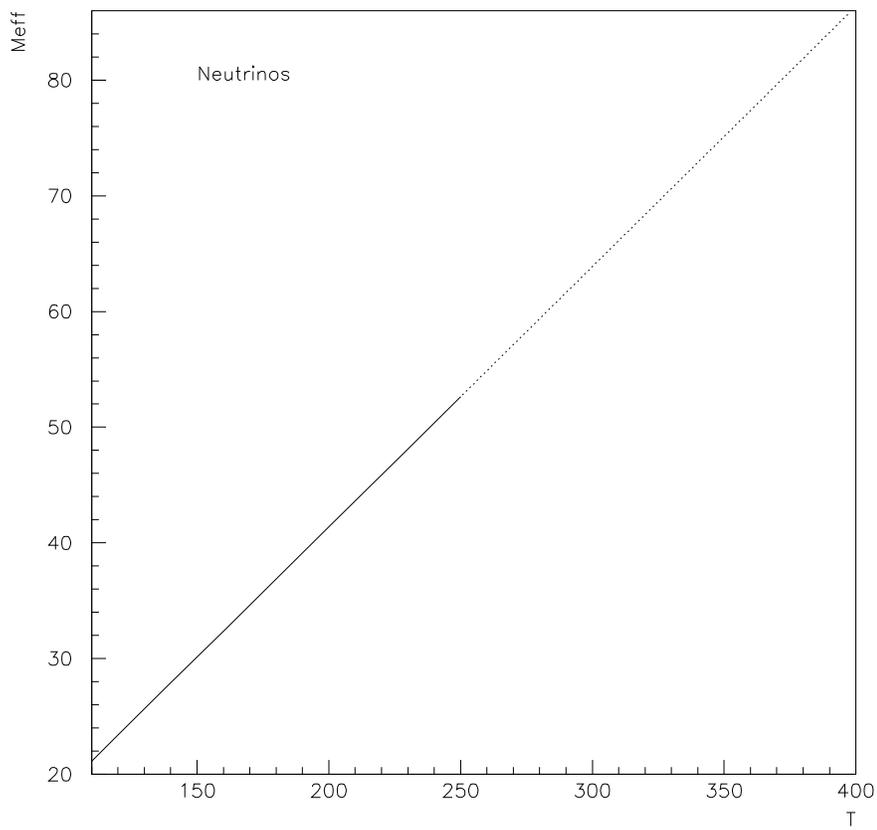}}}\fi
\caption{Variation with temperature of the electron-neutrino
thermal effective masses for the broken (solid curves) and
unbroken (dotted curves) phases}
\label{fig8}
\end{figure}

\newpage

\begin{figure}[9]
\let\picnaturalsize=N
\def\picsize{5in}
\def\picfilename{fig9.ps}
\ifx\nopictures Y\else{\ifx\epsfloaded Y\else\input epsf \fi
\let\epsfloaded=Y
\centerline{\ifx\picnaturalsize N\epsfxsize \picsize\fi
\epsfbox{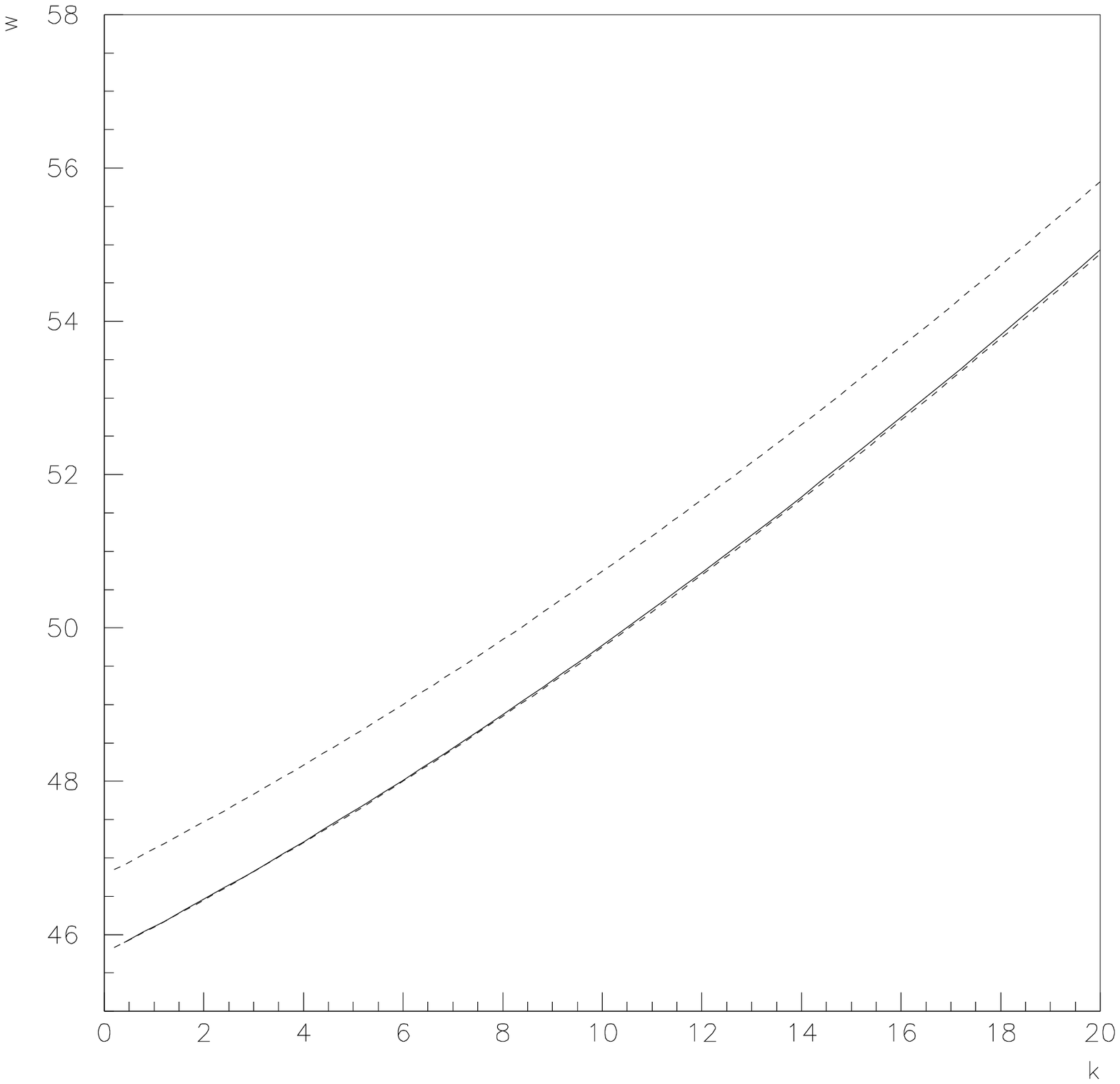}}}\fi
\caption{Normal branch for the massless $b$ quark in QCD computed
for three different cases at $T=100$ GeV. The lower dashed curve
is computed in the Feynman gauge considering only the leading
${\cal O}(T^2)$ contribution to the self-energy (leading curve).
The upper dashed curve corresponds to the results in the Feynman
gauge considering all the contributions to the self energy
(Feynman curve). The solid curve is computed in the Landau
gauge also considering all the contributions (Landau curve). The
leading and Landau curves are superimposed.}
\label{fig9}
\end{figure}

\newpage

\begin{figure}[10]
\let\picnaturalsize=N
\def\picsize{5in}
\def\picfilename{fig10.ps}
\ifx\nopictures Y\else{\ifx\epsfloaded Y\else\input epsf \fi
\let\epsfloaded=Y
\centerline{\ifx\picnaturalsize N\epsfxsize \picsize\fi
\epsfbox{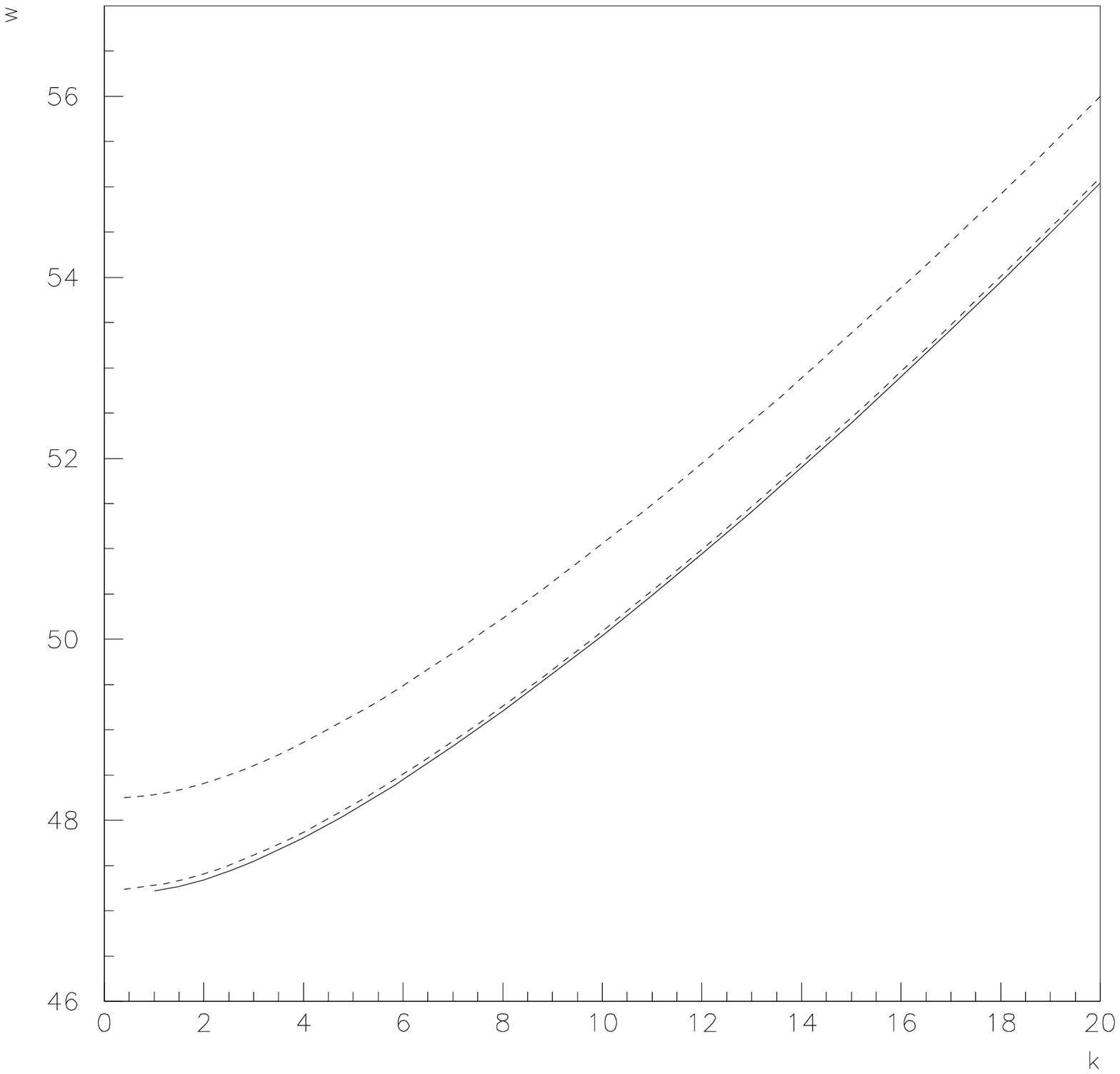}}}\fi
\caption{Normal branch for the massive $b$ quark in QCD computed
for three different cases at $T=100$ GeV. The lower dashed curve
is computed in the Feynman gauge considering only the leading
${\cal O}(T^2)$ contribution to the self-energy (leading curve).
The upper dashed curve corresponds to the results in the Feynman
gauge considering all the contributions to the self-energy
(Feynman curve). The solid curve is computed in the Landau
gauge also considering all the contributions (Landau curve). The
leading and Landau curves are superimposed.}
\label{fig10}
\end{figure}
\end{document}